\documentclass[12pt]{article}          
\input amssym.tex
\def\NB#1\NE{\color{red}#1\color{black}}
\def\OB#1\OE{\color{blue}#1\color{black}}
\def\CB#1||#2\CE{\color{blue}#1\color{red} #2\color{black} }

\usepackage{%
pstricks,%
pst-plot,%
epsf%
}
\usepackage{cite}
\usepackage{graphicx}
\usepackage{array,dcolumn}      
\usepackage{axodraw}
\usepackage{wick}

\newcommand{\hs}{\hspace*{0.5cm}}
\newcommand{\vs}{\vspace*{0.5cm}}
\newcommand{\be}{\begin{equation}}
\newcommand{\ee}{\end{equation}}
\newcommand{\bea}{\begin{eqnarray}}
\newcommand{\eea}{\end{eqnarray}}
\newcommand{\bary}{\begin{array}}
\newcommand{\eary}{\end{array}}
\newcommand{\bit}{\begin{itemize}}
\newcommand{\eit}{\end{itemize}}
\newcommand{\ben}{\begin{enumerate}}
\newcommand{\een}{\end{enumerate}}
\newcommand{\crn}{\nonumber \\}

\newcommand{\al}{\alpha}
\newcommand{\la}{\lambda}
\newcommand{\bet}{\beta}

\newcommand{\fr}{\frac}

\newcommand{\bc}{\begin{center}}
\newcommand{\ec}{\end{center}}

\newcommand{\nn}{\nonumber}

\newcommand{\si}{\sigma}

\begin{document}

 \bc {\Large \bf  General  formula for symmetry factors\\  of Feynman
 diagrams}

\vspace*{1cm}

\textbf{L. T. Hue, H. T. Hung and H. N. Long}

\vspace*{0.5cm}

 \emph{Institute of Physics, VAST, 10 Dao Tan, Ba Dinh,10000 Hanoi, Vietnam }\ec

\begin{abstract}

  General formula for symmetry factors (S-factor) of Feynman diagrams
  containing fields with high spins is derived. We prove that symmetry
  factors of Feynman diagrams of well-known theories do not depend on
  spins of fields. In contributions to S-factors, self-conjugate
  fields and non self-conjugate fields play the same roles as real
  scalar fields and complex scalar fields, respectively. Thus, the
  formula of S-factors for scalar theories --- theories include only
  real and complex scalar fields --- works on all well-known theories
  of fields with high spins.Two interesting consequences deduced from
  our result are : (i) S-factors of all \emph{external connected}
  diagrams consisting of only vertices with  three
  different fields, e.g., spinor QED, are equal to  unity;
  (ii) some diagrams with different topologies can contribute the same
  factor, leading to the result that the inverse
  S-factor for the total contribution is the sum of inverse S-factors,
  i.e., $1/S = \sum_i (1/S_i) $.
\end{abstract}

PACS number(s):  11.15.Bt, 12.39.St.

\noindent

\section{Introduction}
In literature, using perturbation theory and Feynman rules, a
general Green's function of an arbitrary theory can be written in
terms of sum of Feynman diagrams. Each diagram is
associated with a factor known as symmetry factor (S-factor).
There are some ways to calculate this factor such as given in
\cite{an} (for more details, see \cite{AP, HAB, AH, BK}) using
functional derivative method. A computer program \cite{cHA} is
also written based on this method to find out S-factors of
higher-order diagrams from the lower-order ones. Other
 independent approaches base directly on computer programs such as
\cite{qraft,feynart}. S-factors can also be evaluated by using
Wick's theorem, available in many textbooks (see, for example
\cite{cl, ps, ryder, kaku, Greiner}). But the disadvantage of
these books as well as  the  methods mentioned above is none of
them give out any general formulas.  We can see that Ref.\cite{cl}
has an expression for connected diagrams of real scalar theories,
Ref.\cite{Greiner} has some comments about S-factors for scalar
electrodynamics, Ref. \cite{ps} for real scalar $ \phi^4$ and some
particular illustrations in Standard Model.  Especially, the very
detailed investigation into S-factors, which are very close to
weights of Feynman diagrams in $\phi^4$
 theories was presented in  \cite{Kleinert}.
 Refs. \cite{jy,bl} also contain S-factors of some particular
diagrams in QCD.

This paper is the development of \cite{dong} in which we derived
the S-factors for Feynman diagrams of theories with scalar fields.
The definition of S-factor can be found in \cite{ps}. We can
understand this as follows.  Using Wick's theorem for expanding a
Green function one often encounters  many terms whose contractions
are different but contributions are the same. The S-factor is the
number of identical terms which are repeatedly counted.  In
language of Feynman diagram, this factor turns  out to be the
product of total number of (symmetry) permutations of all vertices
and all internal propagators in the diagram, which create new
identical diagrams with factors caused by bubbles. In
Ref.\cite{dong} we have concentrated on two types of fields,
namely real and complex scalar fields, and have noted that the
distinction between these fields is very important because they
contribute different factors to the formula of
S-factor\cite{dong}: \bea S=g 2^\beta 2^d \prod_n (n!)^{\al_n},
\label{scalarSF} \eea where $g$ is the number of interchanges of
vertices leaving the diagram topologically unchanged, $\beta$ is
the number of lines connecting a vertex to itself ($\beta$ is zero
if the field is complex), $d$ is the number of double bubbles, and
$\al_n$ is the number of sets of $n$-identical lines connecting
the same two vertices.

In this paper, by considering some particular cases, we will
indicate precisely that in calculating S-factors, we can classify
all well-known fields into two classes. The first class
comprises self-conjugate fields  for which the particle
is the same as the antiparticle, such as  the real Higgs scalar
$\si$ in the Standard Model, the photon and the $Z$ boson.
We will often refer to this class to be the  real
scalar-like. The second, all non self-conjugate fields --- such as
charged particles --- will be referred to as complex
scalar-like. In analogy with the leptons where $e^-,
\mu^-$ and $\tau^-$ are  called ``particles'' and $e^+,
\mu^+$ and $\tau^+$ are called ``antiparticles'',  hereafter we
 adopt the convention  that the \emph{negative}
electric charged scalar/vector fields (for example $\pi^-, W^-$)
 will be called particles. Keeping these remarks in
mind, we then redefine parameters $g, d, n$ and $\alpha_n$ of
(\ref{scalarSF}) (detailed in the conclusion), then the formula
works on all cases.

One more interesting point we would like to mention about
this paper is  that a simple method of calculating the SF
of a particular Feynman diagram  emerges directly from
the graphical form itself. Especially the $g$-factor, the most
complicated factor appearing in our formula (\ref{scalarSF}) as
well as \cite{cl} and many others textbooks, will be naturally
made clear through our calculation. It relates strictly with
graphical symmetry properties of the diagram.

\hs The outline of our work in this paper is as follows. In the
second section, we recall T-product expansions of interaction
Lagrangians into N-products and introduce a new definition of
vertices and their factors in Feynman diagrams. This helps us
simplify our calculation because, for every interaction
Lagrangian, we will find factors that really contribute to the SFs
and omit other unnecessary factors. The third section is devoted
to spinor QED case, the most simple case that contains spinor
fields.  As mentioned in the abstract, we will show that in the
spinor QED, the S-factor of an arbitrary diagram at any order in
pertubative expansion is always equal to $1$. This is very useful,
for example, in calculating high order QED contributions to lepton
Anomalous Magnetic Moment $(g-2)$ \cite{Tkino}. We will also prove
that spinor fields behave the same as scalar complex fields. In
addition, we will discuss in  detail one interesting way of
practically determining the $g$ factor from the geometric
symmetries of a particular Feynman diagram. In  the next three
sections some particular cases are illustrated to point out that
when calculating S-factor of a diagram, all well-known fields
always belong to one of two classes mentioned above. In the last
section, we will derive the final formula of the S-factor for
general cases. An expression for $g$ factor is also presented in
order to determine it from connected diagrams of the total
diagram. Examples of the S-factors are illustrated in Appendices
\ref{app1} and \ref{app2}.

\section{\label{notaion}Feynman diagrams and symmetry factors}
Let us start using Wick's theorem to expand $T$-products of
interaction  Lagrangians  into sums of $N$-products \cite{an,ps}:
\ben
 \item
Real scalar  $\phi^3$ theory: \bea \mathcal{L}_{int}^r(x) &=&
\fr{\la}{3!} \phi^3(x), \label{Lreal3}\crn \frac{1}{3!}\phi^3(x)
\sim \frac{1}{3!}T\left[\phi^3(x)\right] &=&
N\left[\frac{1}{3!}\phi^3(x)\right] + \frac{3}{3!}
\phi(x)\dot{\Delta}(x) \label{ttichreal3}\eea
  where each  $\dot{\Delta}(x)\equiv \wick{1}{<1\phi(x) >1\phi(x)}$
  corresponds to a
   bubble located at $x$-coordinate in some Feynman diagram.
 \item  Real scalar $\phi^4$ theory:
\bea \mathcal{L}_{int}^r(x) &=&   \fr{\la}{4!} \phi^4(x)
\label{Lreal4}\crn \frac{1}{4!}\phi^4(x) \sim
\frac{1}{4!}T\left[\phi^4(x)\right]& =&
N\left[\frac{1}{4!}\phi^4(x)\right] + \frac{6}{4!}
N\left[\phi^2(x)\right] \dot{\Delta}(x) + \frac{3}{4!}
\dot{\Delta}(x)\dot{\Delta}(x).\crn
 \label{ttichreal4}\eea

\begin{figure}[h]
\begin{picture}(200,100)(-150,80)
\DashLine (-120,130)(-120,150){2}\DashLine
(-140,110)(-120,130){2}\DashLine (-120,130)(-100,110){2}
\DashCArc(-80,140)(10,0,360){2}\DashLine(-80,130)(-80,110){2}
\Text(-100,80)[]{(a) Vertices of $\phi^3$ }
\DashLine(-20,110)(10,140){2}\DashLine(-20,140)(10,110){2}
 \DashLine(15,110)(65,110){2}\DashCArc(40,120)(10,0,360){2}
 \DashCArc(80,120)(10,0,360){2}\DashCArc(80,140)(10,0,360){2}
 \Text(40,80)[]{(b)  Vertices of real $\phi^4$  }
 \DashArrowLine(140,110)(155,125){2}\DashArrowLine(155,125)(170,140){2}
 \DashArrowLine(140,140)(155,125){2}\DashArrowLine(155,125)(170,110){2}
 \DashArrowLine(180,110)(200,110){2}\DashArrowLine(200,110)(220,110){2}
  \DashArrowArc(200,120)(10,270,630){2}
  \DashArrowArc(240,140)(10,270,630){2}
\DashArrowArc(240,120)(10,90,450){2}
 \Text(190,80)[]{(c) Vertices of complex $\varphi^4$  }
 \end{picture}
 \caption[]{ Vertices of  scalar theories }
 \label{vertices}
\end{figure}
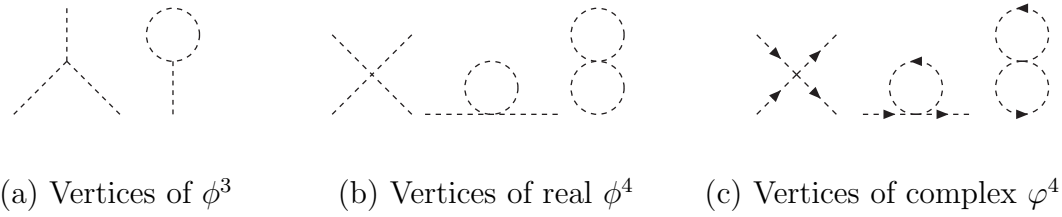

\item   Complex scalar $\varphi^4$ theory:
\bea \mathcal{L}_{int}^c(x) &=&   \fr{\rho}{4}
[\varphi(x)\varphi^*(x)]^2, \crn
\frac{1}{4}[\varphi(x)\varphi(x)^*]^2 \sim
\frac{1}{4}T\left[\varphi(x)\varphi(x)^*\right]^2&=&
N\left[\frac{1}{4}[\varphi(x)\varphi(x)^*]^2\right] + \frac{4}{4}
N\left[\varphi(x)\varphi(x)^*\right] \dot{\Delta}(x)\crn &+&
\frac{2}{4} \dot{\Delta}(x)\dot{\Delta}(x). \label{ttichreac4}\eea
\een
  Each term in right hand sides (RHS) of  (\ref{ttichreal3}), (\ref{ttichreal4})
and (\ref{ttichreac4})   changes into one particular kind
of vertex in the language of Feynman diagram. They are
illustrated in Fig.\ref{vertices}, where  propagators of real
fields are represented as dash lines without directions (arrows),
while complex cases are represented as dash lines with
\emph{directions}. Vertices are different from each others in
numbers of lines and kinds of line they have. This is because
terms in RHSs of \ref{Lreal3}-\ref{ttichreac4} are different in
fields and contractions. Now, for a given interaction Lagrangian,
we can show  exactly all kinds of vertex in the theory. This is
very important
 for us to find out not only $g$ factor relating with
vertices but also contributory factors of different kinds of
vertex to S-factors. Vertices themselves  have well-known factors
as \emph{ vertex factors}, which can be ignored due to the fact
that S-factors are independent on them.

Vertex factors, in scalar theories, are simply $i\lambda$ or
$i\rho$, while in the others such as in scalar electrodynamics or
in quantum chromodynamics, where there exist interactions
containing derivatives, are more complicated. For our method, the
S-factor determined from (\ref{scalarSF}) depends on values of
factors, for example: $ 1/(3!)$ in $\phi^3$, of interacting terms
in Lagrangian. These factors are obtained by taking partial
derivatives of respective interacting terms. Furthermore, we need
to write down each interacting term of the Lagrangian in form of
[vertex factor $\times$ $T$-product], then omit this vertex factor
in our calculation. The symmetry factor now depends only
on $T$-product.

A general Feynman diagram derived from the expansion of a general
Green's function consists of many connected pieces (subdiagrams)
disconnected with each others.  We will call  pieces connected
vacuum diagrams if they have not any external legs and connected
external diagrams if they have at least one external leg [for
example, see Fig.\ref{ad2}(a.10)]. Every connected subdiagram has
its private S-factor. In this work, we concentrate  on only aspect
of S-factor calculation. Other symmetries, such as the charge
conjugation under which diagrams in the QED with odd number of
external photon legs give
vanishing contributions, are outside the scope.\\
\hs Now we turn to theories of fields with high spins.

\section{Symmetry factors in spinor QED}

\hs In spinor Quantum Electrodynamics (QED), the interaction
Lagrangian of one  fermion field $\psi$ is given by \bea
\mathcal{L}_{int}^{QED}(x) &=& e q \overline{\psi}(x)\gamma^\mu
\psi(x)A_{\mu}(x),
 \label{Lqed1}
 \eea where $e$ is the electromagnetic coupling constant, $q$ is the
 electric charge of fermion $\psi$ in units of positron charge and
 $A_\mu(x)$ is the electromagnetic field. For the sake of brevity,
  from now on  we will write $\mathcal{L}(x)$ instead of
 $\mathcal{L}_{int}(x)$.

The above  Lagrangian has only one interacting term with a
vertex factor [$i e q\gamma^\mu$].
 $T$-product expansion  gives:
\bea T\left[\overline{\psi}(x)\gamma^\mu
\psi(x)A_{\mu}(x)\right]&=& N\left[\overline{\psi}(x)\gamma^\mu
\psi(x)A_{\mu}(x)\right]+\wick{2}{<2{\overline{\psi}}(x)\gamma^\mu
>2\psi(x)}A_{\mu}(x)\crn
&=& N\left[\overline{\psi}(x)\gamma^\mu \psi(x)A_{\mu}(x)\right]+i
S^\alpha_\beta(x)(\gamma^\mu)^\beta_\alpha A_{\mu}(x),\label{Td1}
 \eea
 where $ S(x)$ is a fermion bubble.

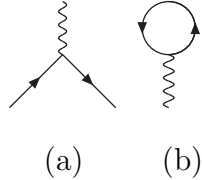
\begin{figure}[h]
\begin{picture}(0,100)(-300,80)

\Photon (-120,130)(-120,150){1.5}{4}\ArrowLine
(-140,110)(-120,130)\ArrowLine (-120,130)(-100,110)
\ArrowArc(-80,140)(10,0,360)\ArrowArc(-80,140)(10,270,90)
\Photon(-80,130)(-80,110){2}{4}
 \Text(-120,90)[]{(a)}
  \Text(-75,90)[]{(b)}
 \end{picture}
 \caption{Vertices of QED  }
 \label{vertQed}
\end{figure}

 The last expression in (\ref{Td1}) has two terms corresponding to two
 kinds of vertices: The first has one photon leg, one
  incoming  and one  outgoing
 electron leg.  The second has one photon leg and one fermion
 bubble. These vertices are illustrated in
 Fig.\ref{vertQed}.

 For simplicity in calculating, let us  denote two terms in the
LHS of (\ref{Td1}) as  follows:
 \bea a_1=N\left[\overline{\psi}(x)\gamma^\mu
\psi(x)A_{\mu}(x)\right],\;\;  a_2=
iS^\alpha_{\beta}(x)(\gamma^\mu)^\beta_\alpha A_{\mu}(x) \equiv i
S(x)\gamma^\mu A_{\mu}(x)\label{T-tichqcd2}\eea Thus, (\ref{Td1})
is rewritten as:
 \bea T[\overline{\psi}(x)\gamma^\mu \psi(x)A_{\mu}(x)]= a_{1}(x)+a_{2}(x) \label{Td2} \eea
The $ n$-point Green's function
 $G^{QED}(x_1,x_2,...,x_n)$ of QED is defined as

 \bea
 G^{QED}(x_1,x_2,...,x_n)&=&
 \sum_{p=0}^\infty\frac{(-i)^p}{p!}\int dy_1 dy_2 ...dy_p\langle0|
 T\left[\phi(x_1)\phi(x_2)...\phi(x_n)\right.\crn
 &\times&\left.\mathcal{L}(y_1)
 \mathcal{L}(y_2)...\mathcal{L}(y_p)\right]|0\rangle,
 \label{GFqed1}\eea
where $\phi(x)$ implies  a spinor $\overline{\psi}(x),\psi(x) $ or
an $A_{\mu}(x)$.
 The $p$th-order term  in this expression is:
 \bea
 G^{QED(p)}(x_1,x_2,...,x_n)&=&
 \frac{(-i)^p}{p!}\int dy_1 dy_2 ...dy_p\langle0|
 T\left[\phi(x_1)\phi(x_2)...\phi(x_n)\right.\crn
 &\times&\left.\mathcal{L}(y_1) \mathcal{L}(y_2)...\mathcal{L}(y_p)\right]|0\rangle
 \label{GFqed2}\eea

 Note that QED has some features different from scalar
 cases.  The Lagrangian of QED contains nonvanishing-spin fields,
 namely half-integer spin fields and spin-1 photon. Spinor fields
 follow anti-communication relations so when positions of these fields
 are changed in a product, a  minus sign will appear. However, it
 does not affect  the S-factors. Furthermore, every
 interacting term always has {\it even } number of spinor fields so
 the value of total product in (\ref{GFqed2}) is unchanged regardless
 positions of these terms. This conclusion is correct for any
 theories. Then the method used in \cite{dong} can again be used as we
 will discuss next.

 In a resulting product [$\mathcal{L}(y_1)
 \mathcal{L}(y_2)...\mathcal{L}(y_p)$] of (\ref{GFqed2}), all terms
 consisting of the same numbers of $a_i$, have the same
 contributions. Then the sum of these terms is presented as a product
 of single symbolic term  multiplied by a factor
 deduced from the multinomial formula: \bea (x_1 +
 x_2 + \cdot \cdot \cdot + x_r)^p & =
 &\sum_{p_1,p_2,...,p_r}\fr{p!}{p_1! p_2! \cdot \cdot \cdot p_r!}
 x_1^{p_1} \cdot \cdot \cdot x_r^{p_r},\label{sf9} \\
 \textrm{with}\hs p_1 + p_2 + p_3 + \cdot \cdot \cdot +p_r & = &
 p. \nn\eea In case of QED, sum of all terms in (\ref{GFqed2}) which
 have $p_1$ factors of $a_1$ and $p_2$ factors of $a_2$ are all
 presented as product of a single term $a_1^{p_1}a_2^{p_2}$
 ($p_1+p_2=p$)  multiplied by a factor: \bea \frac{p!}{p_1!
   p_2!} \label{factor1}\eea

In order to do contractions between internal fields of
$a_1^{p_1}a_2^{p_2}$ and external fields $\phi(x_1)$, $\phi(x_2)$,
$...,\phi(x_n) $, we write the T-product as follows:
 \bea&& \langle0|T[\phi(x_1)\phi(x_2)...\phi(x_n)a_1^{p_1}a_2^{p_2}]|0\rangle\crn
 &=&\langle 0| T[\phi(x_1)\phi(x_2)...\phi(x_n)
 N[\overline{\psi}(y_1)\gamma^{\mu_1}
\psi(y_1)A_{\mu_1}(y_1)]...\label{GFqed3}\\
&\times&N[\overline{\psi}(y_r)\gamma^{\mu_r}
\psi(y_r)A_{\mu_r}(y_r)](-i)S(y_{r+1})\gamma^{\mu_{r+1}}
A_{\mu_{r+1}}(x)...(-i)S(y_{r+s})\gamma^{\mu_{r+s}}
A_{\mu_{r+s}}(x)]|0\rangle\nn \eea

Performing contractions between fields, we rewrite (\ref{GFqed3})
 as a sum of terms containing only contractions.  The terms
with similar contractions lead to the fact that a total
contribution of these terms can be presented as a product of a
term (now corresponding to a Feynman diagram) multiplied
 by a new additional factor.  This new factor contributes to
our S-factor and will be calculated by performing permutations of
propagators and vertices.

Let us concentrate on symmetries of (\ref{GFqed3}) because this
will help us count the number of different terms having the same
contribution. There are factors created by two kinds of
permutation: ($i$) permutation of propagators (lines) in each
vertex and ($ii$) permutation of vertices in one diagram.  For
propagator permutations, first, there is no permutation in the
vertex of type ($a_2$) (figure \ref{vertQed}.b) having only one
leg. Second, according to Wick's theorem, each vertex of type
{($a_1$) has three different fields with possibility of
  contraction with internal fields of other vertices or external
  fields. These three contractions present as three different lines
  (see fig.\ref{vertQed}a). Again, there are no permutations of these
  lines. Therefore the factor caused by propagator permutations in any
  vertex  is $f_1=1$.

  Next, consider symmetries (or equivalences) between vertices.
  Vertices in the same kind, which play the same roles in doing
  contractions, will create distinguishable terms with identical
  contributions. For  QED, there are two types of
  vertex, number of these terms is given by a factor: \bea
  f_2=\frac{p_1 !p_2 !}{g}
\label{factor5}\eea
 Let us explain how to obtain  this
factor. $(p_1!p_2!)$ is the permutation number of vertices ($p_1$
of $a_1$ and $p_2$ of $a_2$) to get new terms, including
repeatedly permutations. Factor $g$ cancels the repeat, i.e.,
permutations are repeatedly counted. It will be more clear when we
discuss directly on Feynman diagrams.

\hs Combining two factors of (\ref{factor1}), (\ref{factor5}) and
the expanding factor ($1/p!$) in (\ref{GFqed2}) we get a total
factor of a particular Feynman diagram appearing in
(\ref{GFqed2}):\bea \frac{1}{S}&=&\frac{1}{p!}\times
\frac{p!}{p_1! p_2!}\times \frac{p_1!p_2!}{g}\crn
&=&\frac{1}{g}\label{gsfqed1}\eea

\hs Hence, in  QED the symmetry factor is only $g$, the factor
belongs to $f_2$ factor. It is more easy to understand $g$ in
language of Feynman diagram. We will see that $g$ is the number of
{\it repeatedly vertex permutations}, i.e., the number of vertex
permutations that creates identical diagrams. Determining $g$ is
rather complicated because we have to make clear relations not
only between vertices themselves but also vertices and
propagators. Fortunately, we can exploit relations between
permutation symmetries and geometrical symmetries of a diagram to
solve this problem. Further, $g$ factor of a general diagram can
be established from $gs$ of connected pieces. Therefore, to
determine $g$ we just consider particular connected subdiagrams
based on geometric symmetries of itself. Let us illustrate this by
some examples in Fig.\ref{vd1}

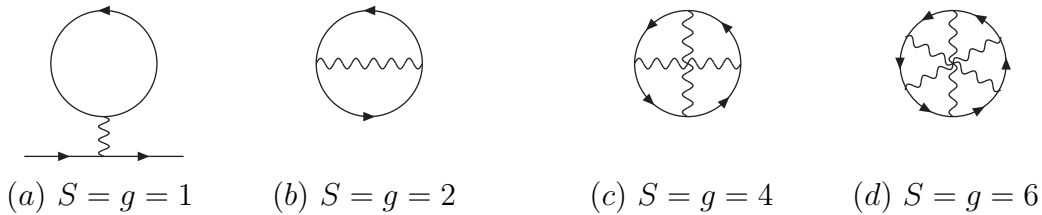
\begin{figure}[h]
\begin{picture}(200,100)(-150,80)

\ArrowArc(-100,150)(20,270,630)
 \Text(-100,100)[]{$(a) \ S=g=1$ }
\Photon(-100,130)(-100,115){2}{3}
 \ArrowLine(-130,115)(-100,115)\ArrowLine(-100,115)(-70,115)

\ArrowArc(0,150)(20,0,180) \ArrowArc(0,150)(20,180,360)
\Photon(-20,150)(20,150){2}{6} \Text(0,100)[]{$(b) \ S=g=2$ }

\ArrowArc(120,150)(20,0,90) \ArrowArc(120,150)(20,90,180)
\ArrowArc(120,150)(20,180,270) \ArrowArc(120,150)(20,270,360)
\Photon(140,150)(100,150){2}{6}
\Photon(120,170)(120,130){2}{6}\Text(120,100)[]{$(c) \ S=g=4$ }

\ArrowArc(220,150)(20,30,90)
\ArrowArc(220,150)(20,90,150)\ArrowArc(220,150)(20,150,210)
\ArrowArc(220,150)(20,210,270) \ArrowArc(220,150)(20,270,330)
\ArrowArc(220,150)(20,330,30) \Photon(220,170)(220,130){1.5}{6}
\Photon(202,140)(238,160){1.5}{6}
\Photon(238,140)(202,160){1.5}{6}\Text(220,100)[]{$(d) \ S=g=6$ }
 \end{picture}
 \caption{Examples of symmetry factors in QED   } \label{vd1}
\end{figure}

\hs  Let us look at  figures \ref{vd1}(b), (c) and (d). Figure (c)
has only rotational symmetries of a square,
 and (d)-a regular hexagon, because  fermion lines have directions. With these
 three diagrams we can rotate (b) an angle $180^0$, (c) three angles $90^0,
180^0, 270^0$ and (c)-$k\times 60^0,k=2,...,5$ to get the same
diagrams as the origins. Clearly, the number of rotative
symmetries (including trivial rotation) is exactly equal to $g$
factor found in \cite{AP}.
 Let us  go to other examples in figure \ref{vd2}. Diagram in figure
 \ref{vd2}a has three identical fermion loops, lying on three vertices of a
 regular triangle, and three photon propagators (no direction).
 Then $g$ of the diagram is $3!=6$, is also equal to six symmetries of a
 regular triangle (two  rotation symmetries, three axial and the
 identical). In Fig.\ref{vd2}b, there is no symmetry because of
 two fixed external propagators. Fig.\ref{vd2}c has three
 connected pieces: two external connected pieces
 do not contribute any factor while the third (connected vacuum piece)
  causes a factor $g =2$ by a $180^0$ rotation.
  From the above  discussion, all S-factors of
  diagrams in the  QED  given in Ref.\cite{AP}, can be derived.
  It is worth noting that in the case of spinor QED, all  connected pieces relating with
  external legs, have $S=1$ because  external legs cancel their geometrical symmetries.

 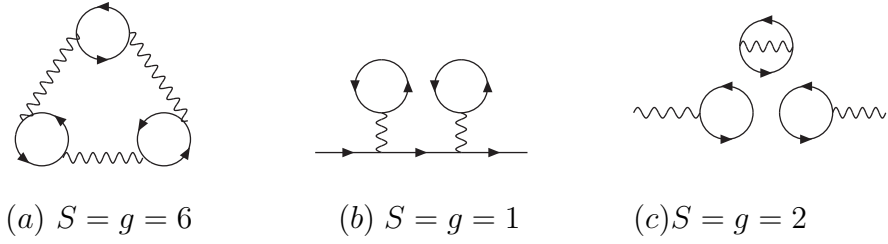
\begin{figure}[h]
\begin{picture}(100,100)(-150,80)

\ArrowArc(-100,170)(10,180,360)\ArrowArc(-100,170)(10,0,180)
\ArrowArc(-123,130)(10,290,170)\ArrowArc(-123,130)(10,170,290)
 \ArrowArc(-77,130)(10,60,240) \ArrowArc(-77,130)(10,240,60)
\Photon(-110,170)(-130,137){2}{8} \Photon(-90,170)(-68,137){2}{8}
\Photon(-85,123)(-115,123){2}{6}
 \Text(-100,100)[]{$(a) \ S=g=6$ }

\ArrowLine(-20,125)(5,125) \ArrowLine(5,125)(35,125)
\ArrowLine(35,125)(60,125)
 \Photon(5,125)(5,140){2}{3}
\Photon(35,125)(35,140){2}{3} \ArrowArc(05,150)(10,-90,90)
\ArrowArc(5,150)(10,90,-90)

   \ArrowArc(35,150)(10,90,-90)
 \ArrowArc(35,150)(10,-90,90)
   \Text(25,100)[]{$(b)\;S=g=1$ }

\Photon(100,140)(125,140){2}{4} \Photon(175,140)(195,140){2}{4}
 \ArrowArc(135,140)(10,0,180)
 \ArrowArc(135,140)(10,-180,0)
 \ArrowArc(165,140)(10,-180,0)
 \ArrowArc(165,140)(10,0,180)
 \ArrowArc(150,165)(10,0,180)
 \ArrowArc(150,165)(10,-180,0)
\Photon(140,165)(160,165){2}{4} \Text(135,100)[]{$(c) S=g=2$ }
 \end{picture}
 \caption{Examples in calculation of symmetry factors }
 \label{vd2}
\end{figure}

One more new important conclusion for this section is:  in our
calculation, fermion fields behave exactly the same as complex
scalar fields, except a minus sign for each closed fermion line.
For the photon $A_\mu$, it has properties of a real scalar field
as we will prove in the next section. Thus $S$ in (\ref{gsfqed1})
is a special case  of formula (\ref{scalarSF}). In appendix
\ref{app1}, S-factors of the  QED up to fourth order are
presented. Ours results are consistent with those in \cite{AP}.

\section{Symmetry factors in scalar Quantum Electrodynamics}

\hs  In scalar  Quantum ElectroDynamics (sQED), the interaction
Lagrangian  consists of both $A_\mu$ and a complex scalar field:
\bea
\mathcal{L}^{sQED}(x)=ieqA^{\mu}(x)[\varphi^*(x)\partial_\mu\varphi(x)
-(\partial_\mu\varphi^*(x))\varphi(x)]+e^2q^2 A_{\mu}(x) A^\mu(x)
\varphi^*(x)\varphi(x),\label{SLqed1}\eea where $q$ is the
electric charge of the complex scalar field $\varphi$.  First, we
pay attention to the term with derivative. This term should be
considered in momentum-space where $\varphi(x)$, $\varphi^*(x)$
and $ A_\mu(x)$ have respective  momenta $p$, $p'$ and $k$. If
$\partial^p_\mu$ denotes that $\partial_\mu$ acts only on field
having momentum  $p$ then we can rewrite:
\be[\varphi^*(x)\partial_\mu\varphi(x)
-(\partial_\mu\varphi^*(x))\varphi(x)]\equiv
(\partial^p-\partial^{p'})_\mu [\varphi^*(x)\varphi(x)]\label{dv
Sqed} \ee

This definition helps us easily write down the vertex factor of
derivative term in  momentum space as  $[ie q(p+p')^\mu ]$.

 \hs $T$-product expansion of the Lagrangian into sum of
$N$-products gives:
 \bea
&&T\left\{ieqA^{\mu}(x)[\partial^p-\partial^{p'}]_\mu
[\varphi^*(x)\varphi(x)]+e^2q^2 A_{\mu}(x) A^\mu(x)
\varphi^*(x)\varphi(x)\right\}\crn&=& ie
q(\partial^p-\partial^{p'})^\mu
A_{\mu}(x)N[\varphi^*(x)\varphi(x)]+ie
q(\partial^p-\partial^{p'})^\mu
A_{\mu}(x)\dot{\Delta}(x)\crn&+&e^2g^{\mu\nu}q^2 N[A_{\mu}(x)
A_\nu(x) \varphi^*(x)\varphi(x)]+(2 e^2 q^2g^{\mu\nu}) \frac{1}{2}
N[A_{\mu}(x)
A_\nu(x)]\dot{\Delta}(x)\crn&+&(2e^2q^2g^{\mu\nu})\frac{1}{2}\dot{\Delta}_{\mu\nu}(x)
N[\varphi^*(x)\varphi(x)]+ (2e^2q^2g^{\mu\nu})
\frac{1}{2}\dot{\Delta}_{\mu\nu}(x)\dot{\Delta}(x)\crn &=& [ie
q(\partial^p-\partial^{p'})^\mu]a_1+[i e
q(\partial^p-\partial^{p'})^\mu]a_2
+(2e^2q^2g^{\mu\nu})\frac{1}{2}a_3\crn&+&(2e^2q^2g^{\mu\nu})\frac{1}{2}
a_4+(2e^2q^2g^{\mu\nu})\frac{1}{2}a_5 +
(2e^2q^2g^{\mu\nu})\frac{1}{2} a_6\label{T-tichsqcd1},\eea where
\bea a_1&=&A_{\mu}(x)N[\varphi^*(x))\varphi(x)]\crn a_2 &=&
A_{\nu}(x) \dot{\Delta}(x)\crn
 a_3&=& N[A_{\nu}(x)
A_\mu(x) \varphi^*(x)\varphi(x)],\crn a_4&=&N[A_{\nu}(x)
A_\mu(x)]\dot{\Delta}(x),\crn a_5&=&\dot{\Delta}_{\mu\nu}(x)
N[\varphi^*(x)\varphi(x)],\crn
a_6&=&\dot{\Delta}_{\mu\nu}(x)\dot{\Delta}(x). \label{ai7}\eea As
before, here we have denoted $\dot{\Delta}_{\mu\nu}(x)\equiv
 \wick{1}{<1A_\mu(x) >1A_\nu(x)}$.
 Let us explain a reason for the factor $\fr 1 2$ associated with
$a_i, i= 3, 4, 5, 6$. As  mentioned in section \ref{notaion}, we
have to write: $[e^2q^2 A_{\mu}(x)  A^\mu(x)
\varphi^*(x)\varphi(x)]= [(2e^2
g^{\mu\nu})]\times[\frac{1}{2}A_{\mu}(x) A_\nu(x)
\varphi^*(x)\varphi(x)]$  because $[(2e^2 g^{\mu\nu})]$ is
well-known vertex factor in literature so the factor $1/2$ (where
2 is  derived by taking derivatives of $[ A_{\mu}(x) A^\mu(x)
\varphi^*(x)\varphi(x)]$ with respect to all fields) is needed for
our method. We note that the photon  $A_\mu$ is real field.
 Factors $ieq(\partial^p-\partial^{p'})^\mu$
and $ (2e^2q^2g^{\mu\nu}$) are vertex factors, we again ignore
them. The vertices in (\ref{ai7}) are illustrated in
Fig.\ref{vertex2}. Applying (\ref{sf9}), each term of the total
Green's function of scalar QED is product of
$[a_1^{p_1}a_2^{p_2}a_3^{p_3}a_4^{p_4}a_5^{p_5}a_6^{p_6}]$ and a
factor $f_1$:
 \bea
&&\frac{p!}{p_1!p_2!p_3!p_4!p_5!p_6!}a_1^{p_1}a_2^{p_2}\left[\frac{a_3}{2}\right]^{p_3}
\left[\frac{a_4}{2}\right]^{p_4}\left[\frac{a_5}{2}\right]^{p_5}\left[\frac{a_6}{2}\right]^{p_6}
=f_1a_1^{p_1}a_2^{p_2}a_3^{p_3}a_4^{p_4}a_5^{p_5}a_6^{p_6},\crn
f_1&=& \frac{p!}{2^{p_3+p_4+p_5+p_6}p_1!p_2!p_3!p_4!p_5!p_6!}.
\label{f1sqed}\eea Now, as an example, we investigate contractions
of a vertex of type $a_3$ with other vertices. This vertex has two
identical lines, so that in some case we can change roles of these
two lines to create  new terms.

\begin{figure}[h]
\begin{picture}(100,100)(-150,80)

\DashArrowLine(-140,150)(-110,150){2}\DashArrowLine(-110,150)(-80,150){2}
\Photon(-110,120)(-110,150){2}{4}\Vertex(-109,150){2}\Text(-110,100)[]{$(a_1)$
}

\Photon(-30,120)(-30,150){2}{4}\DashArrowArc(-30,160)(10,270,90){2}
\DashArrowArc(-30,160)(10,90,270){2}\Vertex(-29,150){2}\Text(-30,100)[]{$(a_2)$
}

\Photon(20,120)(40,140){2}{4}\Photon(40,140)(60,120){2}{4}
\DashArrowLine(20,160)(40,140){2}\DashArrowLine(40,140)(60,160){2}
\Vertex(40,140){2.3}\Text(40,100)[]{$(a_3)$ }

\Photon(88,139)(148,139){2}{8}
\DashArrowArc(120,150)(10,90,270){2}\DashArrowArc(120,150)(10,270,90){2}
\Vertex(120,140){2.3}\Text(125,100)[]{$(a_4)$ }

\DashArrowArc(180,150)(10,90,270){2}\DashArrowArc(180,150)(10,270,90){2}
\PhotonArc(181,129)(10,0,360){1.5}{12}
\Vertex(180,140){2.3}\Text(180,100)[]{$(a_5)$ }

\PhotonArc(239,132)(10,0,360){1}{12}
\DashArrowLine(210,120)(240,120){2}\DashArrowLine(240,120)(270,120){2}
\Vertex(240,120){2.3}\Text(240,100)[]{$(a_6)$ }
 \end{picture}
 \caption{ Vertices of scalar QED } \label{vertex2}
\end{figure}
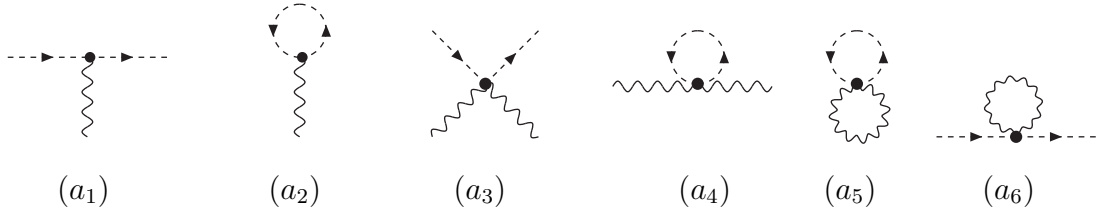

Each vertex of kind $a_1$ or $a_6$ has different fields, $a_5$ has
no relation with other vertices. Vertices $a_3$ and $a_4$, each
has two identical lines. All different contractions of
$a_{1,2,3,4,5,6}$ create a new factor:

\bea f_2= \frac{2^{p_3}2^{p_4}}{\prod_{n}(n!)^{\alpha_n}},
\label{f3sqed} \eea where $n$ and $\alpha_n$ were mentioned in
(\ref{scalarSF}).

\hs Next, similar to QED case, a factor caused from making
contractions of vertices is given by: \bea
\frac{p_1!p_2!p_3!p_4!p_6!}{g'}=f_3 \label{f2sqed},\eea where $
g'$ is number of vertex permutations of types $a_{1, 2,3, 4}$ and
$a_6$ creating identical diagrams. Then, the total factor is: \bea
f=\frac{1}{S}&=&\frac{1}{p!}f_1f_2f_3\crn &=&
\frac{1}{p!}\;\frac{p!}{2^{p_3+p_4+p_5+p_6}p_1!p_2!p_3!p_4!p_5!p_6!}\;
\frac{2^{p_3}2^{p_4}}{\prod_{n}(n!)^{\alpha_n}}\;\frac{p_1!p_2!p_3!p_4!p_6!}{g'}\crn
&=&\frac{1}{g 2^{p_5+p_6}\prod_{n}(n!)^{\alpha_n}}
\label{tfsqed}\;,\eea where $g=g'p_5!$  now is different from
$g'$-permutation number  of all vertices. We note that $g_5!$
related with $a_5$ now is included in $g$. It is easy to realize
that the quantity $\beta$ that appears in Eq. (\ref{scalarSF}) is
given by $p_5+p_6$ for this case.

 From (\ref{tfsqed}) we come to following conclusions:
 \ben
 \item  $a_2$ and $a_4$ do not contribute to $\beta$.
  Remember that this important property of complex scalar fields
   leads to the discrimination
  against real ones.  Also, non self-conjugate  bubble in $a_5$ does not create any new factors.
  $a_5$ in $\beta$ therefore comes from bubbles of $A_\mu$s. As a consequence, we conclude
   that $A_\mu$s play equivalent roles to real scalar fields.
 \item Although this theory contains vertex $a_5$ with two different
    bubbles, the factor $2^d$ does not
   appear. Clearly, $d$ is only the number of vertices with double
   identical bubbles.  This new result is very important for theories
   with many different fields such as [$\phi^2\varphi^2$].  .  \een
   Our formula can be verified by results of
   Ref.\cite{HAB}.

\section{Symmetry factors in  QCD}

\hs The Lagrangian in the  QCD is given by \bea\mathcal{ L}^{QCD}=
\sum_{i=1}^{3}\overline{\psi}(iD_\mu \gamma^\mu-m
)\psi-\frac{1}{4}F^a_{\mu \nu} F^{a\mu\nu},\label{lqcd1}\eea where
$ D_\mu=\partial_\mu-i g_S t_a A^a_\mu$,  $t_a$s are
representation
 matrices of $SU(3)_C$, and $t_a=\frac{\lambda_a}{2}$ for the basic representation,
  $ F^a_{\mu\nu}=\partial_\mu
A^a_\nu-\partial_\nu A^a_\mu+g_S f^{abc}A^b_\mu A^c_\nu $. The
indices $a, b, c=1, 2,..., 8$, $\psi$ has three color components:
$\psi=(\psi^R, \ \psi^G, \ \psi^B)^T$, $A^a_\mu$s are gauge gluon
fields.

\hs Expanding the above  Lagrangian, we have an interaction
lagrangian: \be
 \mathcal{L}_{int}^{QCD}
=g_S\overline{\psi}\gamma^\mu t^a \psi A^a_{\mu }-g_S
f^{abc}(\partial_\mu A^a_{\nu})A^{\mu b}A^{\nu c} -\frac{1}{4}
g^2_S(f^{eab} A^a_{\mu}A^b_\nu)(f^{ecd}A^{\mu c}A^{\nu d})
 \label{Lqcd1}
 \ee We emphasize that QCD is different from  QED, gluon
 gauge fields of QCD $A^a_\mu$s are labeled by a color quantum number
 $a$ and belong to adjoint representation of $SU(3)_C$. But all of
 $A^a_\mu$s are real fields, or self-conjugate
 fields.  Since quarks carry colors so gluons must carry them too
 and physical gauge fields are combinations of $A^a_\mu$s. However,
 due to the assumption that all observed particles only are color
 singlet, we can work with just $A^a_\mu$s \cite{bl}. It is easy to
 see that the first interacting term identifies with T-product.
 Hence, we can write this $T$-product in  terms of sum of
 $N$-products\cite{bl}: \bea T\{\bar{\psi}\gamma^\mu t^a
 A^a_\mu\psi\}&=&N[\bar{\psi}\gamma^\mu t^a A^a_\mu\psi
 ]+N[\wick{2}{<2{\overline{\psi}}_{(x)}\gamma^\mu t^a A^a_\mu
   >2\psi_{(x)}}] \crn&=&[\bar{\psi}\gamma^\mu t^a A^a_\mu\psi
 ]+iS^{i,\alpha}_{k,\beta}(x)(\gamma^\mu)^{\beta}_{\alpha}(t^a)^{k}_{i}
 A^a_\mu,\label{T-tichqcd1}\eea where $\al, \bet$ are Dirac
 indices, and  $i, k$ are $SU(3)_C$ ones.
 For the third term of (\ref{Lqcd1}), firstly we rewrite it in new
 form \cite{bl}: \bea\frac{1}{4} (f^{eab}
 A^a_{\mu}A^b_\nu)(f^{ecd}A^{\mu c}A^{\nu
   d})&=&\left[f^{eab}f^{ecd}(g^{\mu\alpha}g^{\nu\beta}-g^{\mu\beta}g^{\nu\alpha})
 \right.\crn&+&\left.f^{eac}f^{ebd}(g^{\mu\nu}g^{\alpha\beta}-g^{\alpha\nu}g^{\mu\beta})
 \right.\crn&+&\left.f^{ead}f^{ecb}(g^{\mu\alpha}g^{\beta\nu}-g^{\beta\alpha}g^{\mu\nu})
 \right]\times\left[\frac{1}{4!} A^a_{\mu}A^b_\nu A^c_{\alpha
   }A^d_{\beta }\right] \label{term3}\eea Next we choose T-product of
 this term as $\frac{1}{4!}T \left[A^a_{\mu}A^b_\nu A^c_{\alpha
   }A^d_{\beta }\right] $. The rest is vertex factor. Then we will
 get: \bea\frac{1}{4!} T( A^a_{\mu}A^b_\nu
 A^{c}_{\alpha}A^{d}_{\beta})&=&\frac{1}{4!}N[ A^a_{\mu}A^b_\nu
 A^c_{\alpha }A^d_{\beta }]+\frac{6}{4!}\wick{1}{<1A^a_{\mu}>1
   A^b_\nu}N[ A^c_{\alpha }A^d_{\beta }]\crn&+&\frac{3}{4!}\wick{1
   2}{<1A^a_{\mu}>1 A^b_\nu<2 A^c_{\alpha }>2A^d_{\beta}}.
\label{T-tichqcd3}\eea We see that (28) is a part of (27) which
brings out S-factors. This part is almost identical with the
expansion of real scalar theory (\ref{ttichreal4}), except indices
$a$ and $\mu$ of gluon fields. However these indices are quiet.
Four field components $A^a_\mu, A^b_\nu, A^c_\alpha$ and
$A^d_\beta$ are the same after doing contractions to form internal
lines without directions. Hence, we can consider them as four
identical real scalar fields. Thus, the S-factor formula of this
case is also given by the formula (\ref{scalarSF}).

\hs    The second term in (\ref{Lqcd1}) contains derivatives, so
it is easier to
 work  in momentum-space. In  this  space, if we denote
  momentum of $A^a_\nu, A^b_\mu$ and $ A^c_\sigma$ correspond to $p,k$ and $q$
 then we can write $\partial^\alpha A^a_\nu\equiv(\partial^\alpha_p)A^a_\nu$,
 etc..., namely:
 \be f^{abc}(\partial^\mu A^a_{\nu})A^b_{\mu}A^{\nu c}\equiv
  f^{abc}(\partial^\mu_p)( A^a_{\nu}A^b_{\mu b}A^c_{\sigma})g^{\sigma \nu}, \label{gluon3}  \ee
 in sense that $\partial^\mu_p$ does not operate on any fields except
those having momentum $p$.

   Now (\ref{gluon3})  can be rewritten in the form \cite{bl}:
  \be  f^{abc}(\partial^\mu A^a_{\nu})A^b_{\mu}A^{\nu c}=
    f^{abc}[g^{\mu\nu}(\partial_p-\partial_k)^\sigma+g^{\mu\sigma}(\partial_k-\partial_q)^\nu
  +g^{\sigma\nu}(\partial_q-\partial_p)^\mu]\times\left[\frac{1}{6}A^a_\nu A^b_\mu A^c_\sigma\right]
   \label{Ttichqcd2}\ee
Again, the first factor in (\ref{Ttichqcd2}) is the three-gluon
interaction vertex factor in momentum-space and the second, the
T-product, is the same as interacting term of $\varphi^3$ real
scalar theory.

\hs In conclusion for calculating S-factor of QCD, all fermion
fields can be considered as scalar complex fields, and gluon
fields play the roles of real scalar fields. Then, the formula
(\ref{scalarSF}) is applicable to the QCD. Some examples are given
in Fig.\ref{exgluon1}.
 These results also agree with those given in Ref. \cite{ps} and \cite{jy}.
\vs

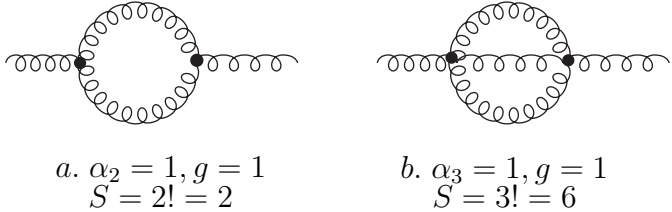
\begin{figure}[ht]
\begin{picture}(400,80)(-150,50)

\Gluon (-140,120)(-112,120){3}{4}\Gluon (-70,120)(-30,120){3}{4}
\GlueArc(-90,120)(20,0,360){3}{20}\Vertex(-112,120){2.3}\Vertex(-68,121){2.5}
\Text(-80,80)[]{$a.\;\alpha_2=1,g=1$ }\Text(-80,70)[]{$S=2!=2$ }

\Gluon (0,120)(30,120){3}{4}\GlueArc(50,120)(20,0,360){3}{20}
\Gluon (70,120)(110,120){3}{4}\Gluon (30,120)(70,120){3}{4}
\Vertex(28,122){2.3}\Vertex(72,121){2.3}
\Text(50,80)[]{$b.\;\alpha_3=1,g=1 $ } \Text(50,70)[]{$S=3!=6 $ }

 \end{picture}
 \caption{ Examples of S-factors in QCD  } \label{exgluon1}
\end{figure}

\section{An example of Standard Model}

\hs Now we turn  to the Standard Model. Let us consider a
particular coupling between $W$ with charged currents: \bea
\mathcal{L}^{CC}_f&=&\frac{g}{\sqrt{2}}\left[W^+_\mu
J^{+\mu}+W^-_\mu J^{-\mu} \right]\crn
 &=& \sum_{i=1}^3 \frac{g}{2\sqrt{2}}\left\{W^{+\mu}\left[\bar{\nu}_i\gamma_\mu(1-\gamma_5)e_i
 +\bar{u_i}\gamma_\mu(1-\gamma_5)d_i\right]\right.\crn
 &+& \left.W^{-\mu}\left[\bar{e_i}\gamma_\mu(1-\gamma_5)\nu_i
 +\bar{d_i}\gamma_\mu(1-\gamma_5)u_i \right]\right\}\label {lag1SM}\eea

This Lagrangian has twelve terms in the same form as $
g/(2\sqrt{2}) \bar{\psi}\gamma_\mu (1-\gamma_5)\psi'W ^\mu$ in
which all terms have the same vertex factor
$[g/(2\sqrt{2})\gamma_\mu (1-\gamma_5)]$. By our choice,
T-products have form $\bar{\psi}\psi'W ^\mu$. It is very simple to
calculate because $T$-product is equal to $N$-product.  In
similarity with  the case of QED, we easily prove that $W$ field
behaves the same way as a complex scalar field.

Our analysis leads to a general principle: interactions such as
given in (\ref{lag1SM}), Yukawa couplings, etc, are similar to
interactions in the spinor QED. Hence we conclude that: S-factors
of all \emph{external connected} (sub-)diagrams containing only
vertices with  three different fields, are equal to
unity. Illustrations can be found in appendix \ref{app2}.
We must remember that $W$ boson is complex scalar-like, in
S-factor calculation, although in diagrams we do not draw its
propagator direction.

It is emphasized that Majorana neutrinos belong to real
scalar-like. For more details, interested readers can find in
Refs.\cite{majorona,pal}.

\section{The vacuum diagrams factorization}

\hs Every Feynman diagram consists of two kinds of well-known
connected pieces, namely external connected and vacuum connected
subdiagrams.  One diagram may  include many identical
vacuum connected pieces. Conversely, all external connected pieces
are different from each others because they connect to different
external legs.  Each piece has its private S-factor which is
independent on the others.

\hs It is interesting that the S-factor of a total diagram
 can be presented  as a product of private S-factors of
connected pieces, that is well known vacuum diagrams
factorization. These private S-factors can be clearly evaluated
from our analysis. If there are $i$ different kinds of connected
piece (easily classified by geometric properties) then we can
label an index $i$ for any thing related with a piece of
the $i$th kind, such as $g_i, \beta_i,d_i,n_i, \alpha_{n_i}$
without losing original meanings of $g,\beta,d,\alpha$. Then, each
connected piece of kind $ith$ contributes a factor $S_i$ to the
total S-factor: \be S_i= g_i 2^{\beta_i} 2^{d_i} \prod_{n_i}
(n_i!)^{\al_{n_i}}, \ee which is the same as (\ref{scalarSF})
except an extra index $i$.  Each set consisting of all $k_i$
indistinguishable pieces (pieces in kind $ith$) causes a factor
$[k_i!(S_i)^{k_i}]$. The total S-factor now is presented as a new
expression: \bea S= \prod_{i} [k_i! (S_i)^{k_i}]=\prod_{i} [k_i!
(g_i)^{i}]\times2^{\sum_{i}k_i\beta_i}\times2^{\sum_{i}k_id_i}\times
\prod_{i}\prod_{n_i} (n_i!)^{\al_{n_i}}\label{SF2} \eea

Comparing with (\ref{scalarSF}) we have
$\beta=\sum_{i}k_i\beta_i$, $d= \sum_{i} k_id_i$ and the factor
$\prod_{i}\prod_{n_i} (n_i!)^{\al_{n_i}}$ can be replaced by $
\prod_{n}(n!)^{\alpha_n}$, where
$\alpha_n=\sum_{i}k_i\alpha_{n_i}$ because
$(n,n_1,n_2,...=1,2,3,...)$ are running indices. Especially, the
relation between $g$ and $g_i$s: \be g=\prod_{i} [k_i!
(g_i)^{k_i}]\label{gfactor1}, \ee can help us practically
calculate $g$ from $g_i$s. The most convenient property
of $g_i$ is that it is equal to the number of graphical symmetry
transformations of a connected piece in the $i$th kind.

 \hs In practice, $\alpha_n, d$ and $\beta$ can directly be deduced from  the
particular  graphical properties of diagram itself. Taking into
account of  (\ref{gfactor1}), we determine $g$ from $g_i$.

\hs To see the vacuum factorization, from (\ref{SF2}), we group
all factors related with connected vacuum pieces (private
S-factors, $S_is$, and factors $k_i!$-rising from $k_i$ identical
pieces) in a single factor called vacuum symmetry factor $S_v$ and
the remaining-external connected ones in another factor $S_c$ ,
then the S-factor of the diagram is divided  into two
factors $S= S_v\times S_c$.  If we sum all of total diagrams in
all orders with their S-factors we will receive results mentioned
in Refs.\cite{cl,ps}.

 \bc
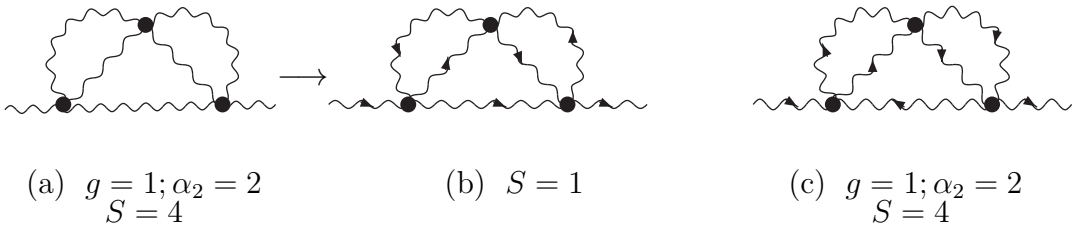
\begin{figure}[ht]
\begin{picture}(400,60)(-150,80)

\Photon(-152,118)(-50,120){1.5}{12}
\Photon(-130,118)(-100,150){1.5}{4}
\Photon(-100,150)(-70,120){1.5}{4} \Vertex(-99,150){3}
\Vertex(-130,120){3}\Vertex(-70,120){3}
\PhotonArc(-115,135)(20,45,225){1.5}{7}
\PhotonArc(-85,135)(20,-45,135){1.5}{7} \Text(-100,90)[]{(a)
$\;g=1;\alpha_2=2$}\Text(-100,80)[]{$ S=4$}

\Text(-40,130)[]{$\longrightarrow$}

\Photon(-30,120)(90,120){1.5}{12} \Photon(0,120)(30,150){1.5}{4}
\Photon(30,150)(60,120){1.5}{4}
\PhotonArc(15,135)(20,45,225){1.5}{7}
\PhotonArc(45,135)(20,-45,135){1.5}{7} \Vertex(31,150){3}
\Vertex(0,120){3}\Vertex(60,120){3} \ArrowLine(62,143)(62,148)
\ArrowLine(14,133)(15,138)\ArrowLine(-5,143)(-3,138)
 \ArrowLine(41,140)(43,135)
  \ArrowLine(-18,121)(-13,119)\Text(40,90)[]{(b) $\;S=1$}
\ArrowLine(32,121)(37,119)\ArrowLine(72,121)(77,119)

\Photon(130,120)(250,120){2}{12} \Photon(160,120)(190,150){2}{4}
\Photon(190,150)(220,120){2}{4}
\PhotonArc(175,135)(20,45,225){2}{7}
\PhotonArc(205,135)(20,-45,135){2}{7} \Vertex(191,150){3}
\Vertex(160,120){3}\Vertex(220,120){3}
\ArrowLine(222,148)(221,143)\ArrowLine(201,141)(201,136)
\ArrowLine(158,138)(156,143)\ArrowLine(175,132)(175,137)
\ArrowLine(142,122)(147,119)\ArrowLine(187,119)(182,122)
\ArrowLine(232,122)(237,119) \Text(190,90)[]{(c) $\;g=1;
\alpha_2=2$ }\Text(190,80)[]{$S=4$}
 \end{picture}
 \caption{   S-factors for diagrams with
different propagator directions }
 \label{Wboson}
\end{figure}
\ec \hs One more remark we point out here: normally, when drawing
a diagram, we just pay attention to directions of momenta while
omitting directions of propagators (for example, $W$
boson). This makes us confused in counting $\beta$ and we may lose
some diagrams because there are diagrams differing \emph{only in
  directions} of propagators. For example, with self-interacting term
of $W$ boson we have a diagram without directions of propagators in
figure \ref{Wboson}.a which has the same S-factor as the one in figure
\ref{Wboson}.c -the figure including charged transition directions of
$W$ (not direction of momentum). But in case of this directional
field, there is another different diagram in figure \ref{Wboson}.b
which does distinguish from the one in figure \ref{Wboson}.c by only
their directions of lines. Both of them have same contributions but
\emph{different} S-factor values. The S-factor now is related with not
only figure \ref{Wboson}.a or \ref{Wboson}.{\bf c} but also with both
of \ref{Wboson}.b and \ref{Wboson}.c \cite {dong}. For simplicity, we
can define a diagram without directions in lines, and call it {\it
  equivalent diagram}. This kind of diagram stands for all diagrams
which  have the same geometrical shape and contribution but are
different in directions of lines. Then the S-factor of a
equivalent diagram is different from usual: it is S-factor for the
total contribution and the inverse of this factor is the sum of
inverse ones of directional diagrams.

To determine S-factor of an equivalent diagram (for example, see,
Fig.\ref{Wboson}a), we have to find out S-factors of all possible
directional diagrams coming from this non-directional diagram, and
denote S-factors $S_1, S_2,...$, respectively. Then the S-factor
 of the equivalent diagram  is
obtained as follows: \bea \frac{1}{S}=\sum_{n}\frac{1}{S_n}.
\label{ctc}\eea

One of our new results that  has never been  mentioned
before: \emph{All well-known formulas for S-factors,
  including  our formula in this paper, only work on directional
  diagrams where all directions of complex scalar-like fields are
  pointed out}. For example, in (\ref{ctc}) our calculation is only
used for $S_n$, not for $S$. From now on, our formula implies
S-factor of $S_n$.

\section{Conclusion}

Based on cases illustrated above, we conclude that our calculation
does not depend on the spins of fields. It  only depends on
whether fields presenting a particle and its anti-particle are
identical or not. In other words, the class of fields is
very important in our calculation of S-factors. We have two
classes of field, real scalar-like and complex scalar-like, as
mentioned in the second section. For practical calculation, in
Table \ref{table1} we list some known fields.
\begin{table}[h]
\caption{
    Classification of fields  }
\begin{center}
\begin{tabular}{|c|c|}
  \hline
  Real scalar-like & Complex scalar-like \\
  \hline
  Real scalar & complex scalar \\
  Photon $A_\mu$ & spinor Dirac field \\
  $Z$ boson & W boson \\
  Gluon &  Ghost \\
  Majarona fields&   \\
  &  \\
  \hline
\end{tabular}
\ec \label {table1}
\end{table}

 \hs Now, as our main result, we introduce a general formula of symmetry factor
 for Feynman diagrams of theories containing
  many different fields with any spin values. Although it has the same form as
  the formula for scalar theories (\ref{scalarSF})
 \bea S=g 2^\beta 2^d \prod_n (n!)^{\al_n}, \label{gSF} \eea
 definitions of $\alpha_n$, $d$ and $\beta$ are  generalized. They are redefined as:
 \begin{itemize}
    \item  $\al_n$ is the number of sets of $n$ identical lines connecting the
 same  pairs of vertices(there may be more than  such one sets in  one vertex pair).
    \item  $d$ is the number of vertices with two {\it identical}
 bubbles.
 \item  $\beta$  is  sum of all self-conjugate bubbles
 coming from  self-conjugate fields ($\beta$ vanishes
  if all fields in the theory belong to non-conjugate fields).
   \item $g$ is the number of vertex permutations keeping
  the diagram topologically unchanged.
  \end{itemize}
    We must emphasize that the most important goal of our work
   is to find out the general definitions of
these parameters in common case. They are more general than
\cite{cl, dong} and others.

The most important thing: formula (\ref{gSF}) is applicable to
diagrams  where all directions of propagators are showed
(although they may not be drawn in diagrams).

We remind  one interesting property of our result: The
diagrams with different topologies can contribute the same, and
the inverse symmetry factor for the total contribution is
therefore the sum of the inverse symmetry ones, i.e., $1/S =
\sum_i (1/S_i) $.

We have showed that  the S-factors of all \emph{external
connected} diagrams containing vertices with three
\emph{different }fields such as interactions in spinor QED, Yukawa
couplings, etc, \emph{are equal to  unity} ($S=1$). This
conclusion is also correct for all diagrams consisting of only
vertices with different legs.

\hs We recall that determining the symmetry factor is important
because it not only is an important component of modern quantum
field theory, but also is used to calculate effective potentials
in higher-dimensional theories and cosmological models. Our
formula works on all of these.

\section*{Acknowledgments}

The authors thank P. V. Dong for useful discussions. Especially,
the authors thank Prof. Palash B. Pal for careful reading and
corrections. This work was supported in part by the National
Foundation for Science and Technology Development (NAFOSTED) under
grant No: 103.01-2011.63.

\appendix
\newpage

\section{\label{app1} Examples of Feynman diagrams in QED up to fourth order.}

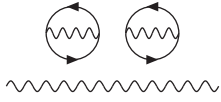
\begin{figure}[ht]
\bc \begin{picture}(400,80)(-150,80)
\ArrowLine(-150,120)(-110,120)\ArrowLine(-110,120)(-70,120)
\Photon(-110,120)(-110,135){2}{3}
\ArrowArc(-110,145)(10,90,-90)\ArrowArc(-110,145)(10,-90,90)
\Text(-100,100)[]{$(a1)\ S=g=1$}
 \ArrowLine(20,120)(100,120)
 \PhotonArc(60,120)(20,0,180){2}{12}
 \Text(50,100)[]{$(b1)\ S=g=1$ }
\ArrowLine(150,120)(230,120)
 \ArrowArc(190,150)(10,0,180)
  \ArrowArc(190,150)(10,-180,0)
  \Photon(180,150)(200,150){2}{4}
  \Text(195,100)[]{$(c1)\ S=g=2$ }
 \end{picture} 
\begin{picture}(300,100)(-80,130)
 \Photon(-130,140)(-110,140){2}{4}
 \Photon(-90,140)(-70,140){2}{4}
 \ArrowArc(-100,140)(10,0,180)
 \ArrowArc(-100,140)(10,180,360)
 \Text(-100,100)[]{$(a2)\  S=g=1 $ }
 \ArrowLine(-30,120)(50,120)
\ArrowArc(-10,150)(10,0,180) \ArrowArc(-10,150)(10,180,360)
\ArrowArc(30,150)(10,0,180)\ArrowArc(30,150)(10,180,360)
\Photon(0,150)(20,150){2}{4}
 \Text(0,100)[]{$(b2)\ S=g=2$ }
\Photon(80,120)(160,120){2}{12}
 \ArrowArc(100,150)(10,0,180)
 \ArrowArc(100,150)(10,180,360)
  \ArrowArc(140,150)(10,0,180)
  \ArrowArc(140,150)(10,180,360)
  \Photon(110,150)(130,150){2}{4}
  \Text(125,100)[]{$(c2)\ S=g=2$ }
\Photon(180,135)(200,135){2}{4} \ArrowArc(210,135)(10,0,180)
 \ArrowArc(210,135)(10,180,360)
 \ArrowArc(235,135)(10,0,180)
 \ArrowArc(235,135)(10,180,360)\Photon(245,135)(265,135){2}{4}
 \Text(230,100)[]{$(d2)\ S=g=1 $ }
 \end{picture}
  \label{ad2}
   \ec
\end{figure}

\begin{figure}[ht]
\bc
\begin{picture}(400,50)(-150,180)
\ArrowLine(-130,120)(-100,120)
 \ArrowLine(-100,120)(-70,120)
 \Photon(-100,120)(-100,130){2}{2}
\ArrowLine(-100,130)(-120,150) \ArrowLine(-80,150)(-100,130)
\ArrowLine(-120,150)(-80,150)
\PhotonArc(-100,150)(20,0,180){2}{12}
 \Text(-100,100)[]{$(a3)\ S=g=1 $ }

  \ArrowLine(15,120)(35,120)
  \ArrowLine(35,120)(65,120)
  \ArrowLine(65,120)(80,120)
  \ArrowLine(80,120)(110,120)
  \ArrowLine(110,120)(130,120)
\PhotonArc(50,120)(15,0,180){2}{8}
\PhotonArc(95,120)(15,0,180){2}{8}
 \Text(60,100)[]{$(b3)\ S=g=1$ }

\ArrowLine(150,120)(175,120) \ArrowLine(175,120)(205,120)
\ArrowLine(205,120)(230,120)
 \ArrowArc(190,160)(10,0,180)
 \ArrowArc(190,160)(10,180,360)
\Photon(180,160)(200,160){2}{4}
  \PhotonArc(190,120)(15,0,180){2}{8}
  \Text(195,100)[]{$(c3)\ S=g=1/2$ }
 \end{picture}
 \begin{picture}(400,50)(-150,230)
\ArrowLine(-150,120)(-120,120)
 \ArrowLine(-120,120)(-100,120)
  \ArrowLine(-100,120)(-70,120)
\Photon(-100,120)(-100,150){2}{6}
 \Photon(-120,120)(-120,150){2}{6}
\ArrowArc(-110,150)(10,0,180) \ArrowArc(-110,150)(10,180,360)
 \Text(-100,100)[]{$(a4)\ S=g=1 $ }

  \ArrowLine(20,120)(100,120)
  \ArrowLine(50,140)(50,170)
  \ArrowLine(50,170)(70,170)
  \ArrowLine(70,170)(70,140)
  \ArrowLine(70,140)(50,140)
  \PhotonArc(50,155)(15,90,270){2}{8}
\PhotonArc(70,155)(15,270,90){2}{8}
 \Text(50,100)[]{$(b4)\ S=g=2$ }

\ArrowLine(150,120)(230,120)
 \ArrowArc(175,160)(10,0,180)
 \ArrowArc(175,160)(10,180,360)
\Photon(165,160)(185,160){2}{4} \ArrowArc(205,160)(10,0,180)
 \ArrowArc(205,160)(10,180,360)
\Photon(195,160)(215,160){2}{4}
  \Text(185,100)[]{$(c4)\ g_1=2;\;S= 2! (g_1)^2=8$ }
 \end{picture}
\ec
\begin{picture}(400,50)(-150,280)
\Photon(-150,120)(-130,120){2}{4}
 \Photon(-90,120)(-70,120){2}{4}
\ArrowLine(-130,120)(-120,120)
 \ArrowLine(-100,120)(-90,120)
 \ArrowLine(-120,120)(-100,120)
  \ArrowArc(-110,120)(20,0,180) \PhotonArc(-110,120)(10,180,0){2}{6}
 \Text(-100,100)[]{$(a5)\ S=g=1 $ }

\Photon(20,120)(40,120){2}{4} \Photon(80,120)(100,120){2}{4}
  \ArrowLine(40,120)(60,120)
  \ArrowLine(60,120)(80,120)
  \ArrowLine(40,120)(60,160)
  \ArrowLine(60,160)(80,120)
 \Photon(60,120)(60,160){2}{8}
 \Text(50,100)[]{$(b5)\ S=g=1$ }

\Photon(150,140)(165,140){2}{3}
\Photon(215,140)(230,140){2}{3}\Photon(185,140)(195,140){2}{2}
 \ArrowArc(175,140)(10,0,180)
 \ArrowArc(175,140)(10,180,360)
 \ArrowArc(205,140)(10,0,180)
 \ArrowArc(205,140)(10,180,360)
  \Text(195,100)[]{$(c5)\ S=g=1$ }
 \end{picture}
\end{figure}

\newpage

\begin{figure}[ht]
\bc
\begin{picture}(400,50)(-150,140)
\Photon(-150,120)(-70,120){2}{12}
 \ArrowArc(-125,140)(10,0,180)
 \ArrowArc(-125,140)(10,180,360)
\Photon(-135,140)(-115,140){2}{4} \ArrowArc(-95,140)(10,0,180)
 \ArrowArc(-95,140)(10,180,360)
\Photon(-105,140)(-85,140){2}{4}
  \Text(-125,100)[]{$(a.6)\; g_1=2,k_1=2 $ }
  \Text(-125,80)[]{$\  S=g=2!(2)^2=8 $ }

   \ArrowLine(20,120)(100,120)
  \ArrowArc(45,150)(10,0,180)
 \ArrowArc(45,150)(10,180,360)
 \Photon(35,150)(55,150){2}{4}
 \Photon(45,120)(45,140){2}{3}
 \ArrowArc(75,150)(10,0,180)
 \ArrowArc(75,150)(10,180,360)
 \Photon(65,150)(85,150){2}{4}
 \Text(50,100)[]{$(b.6)\  S=g=2$ }

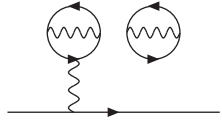
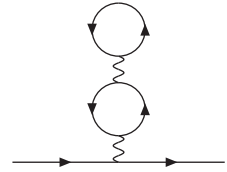
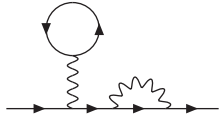
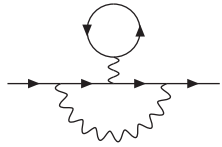
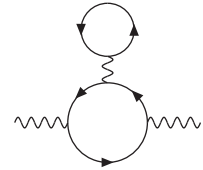
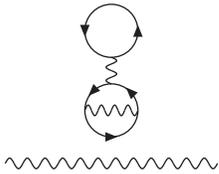
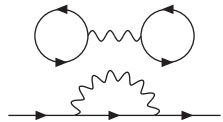
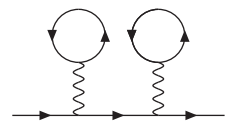
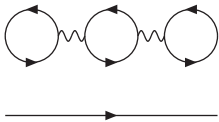
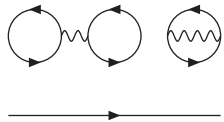
\ArrowLine(150,120)(190,120) \ArrowLine(190,120)(230,120)
\ArrowArc(190,140)(10,90,-90)
 \ArrowArc(190,140)(10,-90,90)
 \Photon(190,120)(190,130){2}{2}
 \Photon(190,150)(190,160){2}{2}
 \ArrowArc(190,170)(10,90,-90)
 \ArrowArc(190,170)(10,-90,90)
   \Text(195,100)[]{$(c.6)\ S=g=1$ }
 \end{picture}
 \begin{picture}(400,80)(-150,220)

\ArrowLine(-150,120)(-125,120) \ArrowLine(-125,120)(-110,120)
\ArrowLine(-110,120)(-90,120) \ArrowLine(-90,120)(-70,120)
 \ArrowArc(-125,150)(10,90,-90)
 \ArrowArc(-125,150)(10,-90,90)
\Photon(-125,120)(-125,140){2}{4}
\PhotonArc(-100,120)(10,0,180){2}{6}
  \Text(-125,100)[]{$(a.7)\ S=g=1$ }

   \ArrowLine(20,140)(40,140)
   \ArrowLine(40,140)(60,140)
   \ArrowLine(60,140)(80,140)
   \ArrowLine(80,140)(100,140)
  \ArrowArc(60,160)(10,90,-90)
 \ArrowArc(60,160)(10,-90,90)
 \Photon(60,140)(60,150){2}{2}
 \PhotonArc(60,140)(20,180,0){2}{10}
   \Text(50,100)[]{$(b.7)\ S=g=1$ }

\Photon(155,140)(175,140){2}{4} \Photon(205,140)(225,140){2}{4}
\ArrowArc(190,140)(15,-180,0) \ArrowArc(190,140)(15,90,180)
 \ArrowArc(190,140)(15,0,90)
 \Photon(190,155)(190,165){2}{2}
  \ArrowArc(190,175)(10,90,-90)
 \ArrowArc(190,175)(10,-90,90)
   \Text(195,100)[]{$(c.7)\ S=g=1$ }
 \end{picture}
 \ec
 \begin{picture}(400,80)(-150,300)

\Photon(-150,120)(-70,120){2}{12}
 \ArrowArc(-110,140)(10,0,90)
 \ArrowArc(-110,140)(10,90,180)
 \ArrowArc(-110,140)(10,-180,0)
\Photon(-120,140)(-100,140){2}{4} \ArrowArc(-110,170)(10,90,-90)
\ArrowArc(-110,170)(10,-90,90) \Photon(-110,150)(-110,160){2}{2}
  \Text(-135,100)[]{$(a.8)\ S=g=1$ }

   \ArrowLine(20,120)(45,120)
   \ArrowLine(45,120)(75,120)
   \ArrowLine(75,120)(100,120)
   \ArrowArc(40,150)(10,0,180)
   \ArrowArc(40,150)(10,-180,0)
   \Photon(50,150)(70,150){2}{3}
   \ArrowArc(80,150)(10,0,180)
 \ArrowArc(80,150)(10,-180,0)
 \PhotonArc(60,120)(15,0,180){2}{8}
    \Text(50,100)[]{$(b.8)\ S=g=2$ }

\ArrowLine(150,120)(175,120) \ArrowLine(175,120)(205,120)
\ArrowLine(205,120)(230,120) \Photon(175,120)(175,140){2}{4}
\Photon(205,120)(205,140){2}{4} \ArrowArc(175,150)(10,-90,90)
\ArrowArc(175,150)(10,90,-90)
 \ArrowArc(205,150)(10,90,-90)
   \ArrowArc(205,150)(10,90,-90)
 \ArrowArc(205,150)(10,-90,90)
   \Text(195,100)[]{$(c.8)\ S=g=1$ }
 \end{picture}
 \begin{picture}(400,80)(-150,380)

\ArrowLine(-150,120)(-70,120)
 \ArrowArc(-110,150)(10,0,180)
 \ArrowArc(-110,150)(10,-180,0)
 \ArrowArc(-140,150)(10,-180,0)
 \ArrowArc(-140,150)(10,0,180)
 \ArrowArc(-80,150)(10,0,180)
 \ArrowArc(-80,150)(10,-180,0)
\Photon(-130,150)(-120,150){2}{2} \Photon(-100,150)(-90,150){2}{2}
  \Text(-135,100)[]{$(a.9)\ S=g=2$ }

   \ArrowLine(20,120)(100,120)
   \ArrowArc(30,150)(10,0,180)
   \ArrowArc(30,150)(10,-180,0)
   \ArrowArc(60,150)(10,0,180)
   \ArrowArc(60,150)(10,-180,0)
   \ArrowArc(90,150)(10,0,180)
   \ArrowArc(90,150)(10,-180,0)
   \Photon(80,150)(100,150){2}{4}
   \Photon(40,150)(50,150){2}{2}
  \Text(50,100)[]{$(b.9) \; g_1=g_2=2;k_1=k_2=1$ }
\Text(50,80)[]{$ S=g=g_1g_2=4$ } \Photon(150,140)(180,140){2}{6}
\Photon(200,140)(230,140){2}{6} \ArrowArc(190,140)(10,0,180)
\ArrowArc(190,140)(10,-180,0) \ArrowArc(170,170)(10,0,180)
\ArrowArc(170,170)(10,-180,0) \ArrowArc(210,170)(10,0,180)
\ArrowArc(210,170)(10,-180,0) \Photon(180,170)(200,170){2}{4}
\Text(195,100)[]{$(c.9)\ S=g=2$ }
 \end{picture}
\end{figure}
\newpage

\begin{figure}[ht]
\bc
\begin{picture}(400,100)(-150,80)

\Photon(-150,140)(-135,140){2}{3} \Photon(-85,140)(-70,140){2}{3}
 \ArrowArc(-125,140)(10,0,180)
 \ArrowArc(-125,140)(10,-180,0)
 \ArrowArc(-95,140)(10,-180,0)
 \ArrowArc(-95,140)(10,0,180)
 \ArrowArc(-110,160)(10,0,180)
 \ArrowArc(-110,160)(10,-180,0)
\Photon(-120,160)(-100,160){2}{4} \Text(-135,100)[]{$(a.10)\
S=g=2$ }

   \Photon(20,150)(30,150){2}{2}
   \ArrowArc(40,150)(10,0,180)
   \ArrowArc(40,150)(10,-180,0)
   \ArrowArc(70,150)(10,0,180)
   \ArrowArc(70,150)(10,-180,0)
   \ArrowArc(100,150)(10,0,180)
   \ArrowArc(100,150)(10,-180,0)
   \Photon(110,150)(120,150){2}{2}
   \Photon(50,150)(60,150){2}{2}
  \Text(50,100)[]{$(b.10)\  S=g=1$ }

\Photon(150,120)(230,120){2}{12} \Photon(150,140)(170,140){2}{3}
\ArrowArc(160,140)(10,0,180) \ArrowArc(160,140)(10,-180,0)
\ArrowArc(185,140)(10,0,180) \ArrowArc(185,140)(10,-180,0)
\ArrowArc(220,140)(10,0,180) \ArrowArc(220,140)(10,-180,0)
\Photon(195,140)(210,140){2}{2} \Text(195,100)[]{$(c.10)\ S=g=4$ }
 \end{picture}

\begin{picture}(400,80)(-150,160)

\Photon(-150,120)(-70,120){2}{12} \ArrowArc(-110,150)(10,0,180)
 \ArrowArc(-110,150)(10,-180,0)
 \ArrowArc(-140,150)(10,-180,0)
 \ArrowArc(-140,150)(10,0,180)
 \ArrowArc(-80,150)(10,0,180)
 \ArrowArc(-80,150)(10,-180,0)
\Photon(-130,150)(-120,150){2}{2} \Photon(-100,150)(-90,150){2}{2}
  \Text(-135,100)[]{$(a.11)\ S=g=2$ }

\ArrowLine(20,120)(35,120) \ArrowLine(35,120)(100,120)
\Photon(35,120)(35,140){2}{4} \ArrowArc(35,150)(10,90,-90)
\ArrowArc(35,150)(10,-90,90) \ArrowArc(60,150)(10,0,180)
\ArrowArc(60,150)(10,-180,0) \ArrowArc(90,150)(10,0,180)
\ArrowArc(90,150)(10,-180,0) \Photon(70,150)(80,150){2}{2}
\Text(50,100)[]{$(b.11)\ S=g=2$ }

\ArrowLine(150,120)(245,120) \Photon(170,140)(180,140){2}{2}
\ArrowArc(160,140)(10,0,180) \ArrowArc(160,140)(10,-180,0)
\ArrowArc(190,140)(10,0,180) \ArrowArc(190,140)(10,-180,0)
\ArrowArc(215,140)(10,0,180) \ArrowArc(215,140)(10,-180,0)
\ArrowArc(245,140)(10,0,180) \ArrowArc(245,140)(10,-180,0)
\Photon(225,140)(235,140){2}{2} \Text(185,100)[]{$(c.11)\
g_1=2,k_1=2$ } \Text(185,80)[]{$ S=g=2!2^2=8$ }
 \end{picture}

 \begin{picture}(400,80)(-150,240)

\ArrowLine(-150,120)(-130,120) \ArrowLine(-90,120)(-70,120)
\ArrowLine(-130,120)(-120,120) \ArrowLine(-120,120)(-100,120)
\ArrowLine(-100,120)(-90,120) \PhotonArc(-110,120)(10,0,180){2}{6}
\PhotonArc(-110,120)(20,0,180){2}{10} \Text(-130,100)[]{$(a.12)\
S=g=1$ }

\ArrowLine(15,150)(35,150) \ArrowLine(85,150)(105,150)
\ArrowLine(35,150)(45,150) \ArrowLine(45,150)(75,150)
\ArrowLine(75,150)(85,150) \ArrowLine(85,150)(100,150)
\PhotonArc(65,150)(20,0,180){2}{10}
\PhotonArc(55,150)(20,-180,0){2}{10} \Text(50,100)[]{$(b.12)\
S=g=1$ }

\ArrowLine(150,120)(170,120) \ArrowLine(170,120)(190,120)
\ArrowLine(190,120)(210,120) \ArrowLine(210,120)(230,120)
\PhotonArc(180,120)(10,0,180){2}{6} \ArrowArc(210,145)(10,-90,90)
\ArrowArc(210,145)(10,90,-90) \Photon(210,120)(210,135){2}{3}
\Text(185,100)[]{$(c.12)\ S=g=1$ }
 \end{picture}

 \begin{picture}(400,80)(-120,320)

\Photon(-150,140)(-130,140){2}{4} \ArrowArc(-120,140)(10,0,180)
 \ArrowArc(-120,140)(10,-180,0)
 \Photon(-85,140)(-65,140){2}{4}
 \ArrowArc(-95,140)(10,-180,0)
 \ArrowArc(-95,140)(10,0,180)
 \ArrowArc(-130,170)(10,0,180)
 \ArrowArc(-130,170)(10,-180,0)
 \ArrowArc(-90,170)(10,0,180)
 \ArrowArc(-90,170)(10,-180,0)
\Photon(-120,170)(-100,170){2}{4}
  \Text(-135,100)[]{$(a.13)\ S=g=2$ }

\Photon(-30,120)(55,120){2}{13} \ArrowArc(-20,150)(10,0,180)
\ArrowArc(-20,150)(10,-180,0) \ArrowArc(10,150)(10,0,180)
\ArrowArc(10,150)(10,-180,0) \Photon(-10,150)(0,150){2}{2}
\ArrowArc(35,150)(10,0,180) \ArrowArc(35,150)(10,-180,0)
\ArrowArc(65,150)(10,0,180) \ArrowArc(65,150)(10,-180,0)
\Photon(45,150)(55,150){2}{2} \Text(0,100)[]{$(b.13)\ S=g=8$ }

\ArrowLine(100,120)(180,120) \ArrowArc(140,135)(10,-90,90)
\ArrowArc(140,135)(10,90,-90) \Photon(140,145)(140,155){2}{2}
\ArrowLine(140,155)(125,175) \ArrowLine(125,175)(155,175)
\ArrowLine(155,175)(140,155) \PhotonArc(140,175)(15,0,180){2}{8}
\Text(135,100)[]{$(c.13)\ S=g=1$ }

\Photon(200,150)(215,150){2}{2}\ArrowLine(215,150)(235,165)
\ArrowLine(235,165)(235,135) \ArrowLine(235,135)(215,150)
\PhotonArc(235,150)(15,-90,90){2}{8}
\Photon(275,150)(290,150){2}{2} \ArrowArc(265,150)(10,0,180)
\ArrowArc(265,150)(10,-180,0) \Text(240,100)[]{$(d.13)\ S=g=1$ }
 \end{picture}
\ec
\end{figure}

\newpage
\section{\label{app2} Examples of Feynman diagrams in SM up to tenth order: $\mu^-
\rightarrow \nu_\mu + e^- + \widetilde{\nu_e}$.} This case we must
remember that all W-boson lines are directional, thought we don't
draw.
\begin{figure}[ht]
\bc
\begin{picture}(400,90)(-150,100)

\ArrowLine(-150,120)(-130,120) \ArrowLine(-130,120)(-110,120)
\ArrowLine(-110,120)(-90,120) \ArrowLine(-90,120)(-50,120)
\Photon(-90,120)(-90,145){2}{4} \ArrowLine(-90,145)(-50,145)
\ArrowLine(-50,185)(-90,145) \PhotonArc(-120,120)(10,0,180){2}{5}
\Text(-150,110)[]{$\mu$} \Text(-40,110)[]{$\nu_\mu$}
\Text(-50,190)[]{$\widetilde{\nu_e}$} \Text(-40,145)[]{$e$}
\Text(-100,100)[]{$(a.14)\ S=1$}

\ArrowLine(0,120)(40,120) \ArrowLine(40,120)(60,120)
\ArrowLine(60,120)(80,120) \ArrowLine(80,120)(100,120)
\Photon(40,120)(40,145){2}{4} \ArrowLine(40,145)(100,145)
\ArrowLine(100,185)(40,145) \PhotonArc(70,120)(10,0,180){2}{5}
\Text(0,110)[]{$\mu$} \Text(110,110)[]{$\nu_\mu$}
\Text(90,190)[]{$\widetilde{\nu_e}$} \Text(110,145)[]{$e$}
\Text(50,100)[]{$(b.14)\ S=1$ }

\ArrowLine(140,120)(190,120) \ArrowLine(190,120)(240,120)
\ArrowArc(190,140)(10,-90,90) \ArrowArc(190,140)(10,90,-90)
\Photon(190,120)(190,130){2}{2} \Photon(190,150)(190,160){2}{2}
\ArrowLine(190,160)(240,160) \ArrowLine(230,185)(190,160)
\Text(140,110)[]{$\mu$} \Text(210,140)[]{$\mu$}
\Text(250,110)[]{$\nu_\mu$} \Text(240,190)[]{$\widetilde{\nu_e}$}
\Text(250,160)[]{$e$} \Text(200,100)[]{$(c.14)\  S =1$ }
 \end{picture}
 \begin{picture}(400,80)(-150,160)
\ArrowLine(-150,120)(-130,120) \ArrowLine(-130,120)(-70,120)
\ArrowLine(-130,140)(-110,140) \ArrowLine(-110,140)(-90,140)
\Photon(-130,120)(-130,140){2}{3} \ArrowLine(-90,140)(-70,140)
\ArrowLine(-70,155)(-130,140)
\PhotonArc(-100,140)(10,-180,0){2}{6} \Text(-150,110)[]{$\mu$}
\Text(-60,110)[]{$\nu_\mu$} \Text(-60,160)[]{$\widetilde{\nu_e}$}
\Text(-60,140)[]{$e$} \Text(-100,100)[]{$(a.15)\ S =1$}

\ArrowLine(0,120)(30,120) \ArrowLine(30,120)(70,120)

\Photon(30,120)(30,140){2}{3} \ArrowLine(30,140)(70,140)
\ArrowLine(70,180)(30,140) \PhotonArc(50,160)(10,-135,45){2}{5}
\Text(0,110)[]{$\mu$} \Text(80,110)[]{$\nu_\mu$}
\Text(80,170)[]{$\widetilde{\nu_e}$} \Text(80,140)[]{$e$}
\Text(40,100)[]{$(b.15)\ S=1$ }

\ArrowLine(140,120)(190,120) \ArrowLine(190,120)(240,120)
\ArrowArc(190,140)(10,-90,90) \ArrowArc(190,140)(10,90,-90)
\Photon(190,120)(190,130){2}{2} \Photon(190,150)(190,160){2}{2}
\ArrowLine(190,160)(240,160) \ArrowLine(230,185)(190,160)
\Text(140,110)[]{$\mu$} \Text(250,110)[]{$\nu_\mu$}
\Text(240,190)[]{$\widetilde{\nu_e}$} \Text(210,140)[]{$e$}
\Text(250,160)[]{$e$} \Text(200,100)[]{$(c.15)\ S=1$ }
 \end{picture}

\begin{picture}(400,80)(-150,200)
\ArrowLine(-150,120)(-135,120) \ArrowLine(-135,120)(-115,120)
\ArrowLine(-115,120)(-100,120) \ArrowLine(-100,120)(-85,120)
\ArrowLine(-85,120)(-65,120) \ArrowLine(-65,120)(-50,120)
\Photon(-100,120)(-100,145){2}{4} \ArrowLine(-100,145)(-50,145)
\ArrowLine(-50,185)(-100,145) \PhotonArc(-125,120)(10,0,180){2}{6}
\PhotonArc(-75,120)(10,0,180){2}{6} \Text(-150,110)[]{$\mu$}
\Text(-40,110)[]{$\nu_\mu$} \Text(-50,190)[]{$\widetilde{\nu_e}$}
\Text(-40,145)[]{$e$} \Text(-100,100)[]{$(a.16)\ S=1$}

\ArrowLine(0,120)(40,120) \ArrowLine(40,120)(60,120)
\ArrowLine(60,120)(80,120) \ArrowLine(80,120)(100,120)
\Photon(40,120)(40,130){2}{2} \Photon(40,150)(40,160){2}{2}
\ArrowArc(40,140)(10,-90,90) \ArrowArc(40,140)(10,90,-90)
\ArrowLine(40,160)(100,160) \ArrowLine(100,185)(40,160)
\PhotonArc(70,120)(10,0,180){2}{5} \Text(0,110)[]{$\mu$}
\Text(110,110)[]{$\nu_\mu$} \Text(90,190)[]{$\widetilde{\nu_e}$}
\Text(110,160)[]{$e$} \Text(50,100)[]{$(b.16)\ S=1$ }
\Text(57,140)[]{$e$}

\ArrowLine(130,120)(150,120) \ArrowLine(150,120)(170,120)
\PhotonArc(160,120)(10,0,180){2}{5} \ArrowLine(170,120)(190,120)
\ArrowLine(190,120)(240,120) \ArrowArc(190,140)(10,-90,90)
\ArrowArc(190,140)(10,90,-90) \Photon(190,120)(190,130){2}{2}
\Photon(190,150)(190,160){2}{2} \ArrowLine(190,160)(240,160)
\ArrowLine(230,185)(190,160)
\Text(140,110)[]{$\mu$}\Text(210,140)[]{$\mu$}
\Text(240,110)[]{$\nu_\mu$} \Text(240,190)[]{$\widetilde{\nu_e}$}
\Text(250,160)[]{$e$} \Text(200,100)[]{$(c.16)\ S=1$ }
 \end{picture}

\begin{picture}(400,90)(-150,230)
\ArrowLine(-150,120)(-70,120) \ArrowLine(-70,120)(-50,120)
\ArrowArc(-125,140)(10,0,180) \ArrowArc(-125,140)(10,-180,0)
\Photon(-135,140)(-115,140){2}{3} \ArrowArc(-95,140)(10,0,180)
\ArrowArc(-95,140)(10,-180,0) \Photon(-105,140)(-85,140){2}{3}
\Photon(-70,120)(-70,145){2}{4} \ArrowLine(-70,145)(-50,145)
\ArrowLine(-50,165)(-70,145) \Text(-95,160)[]{$\mu$}
\Text(-125,160)[]{$\mu$} \Text(-150,110)[]{$\mu$}
\Text(-50,110)[]{$\nu_\mu$} \Text(-50,170)[]{$\widetilde{\nu_e}$}
\Text(-40,145)[]{$e$} \Text(-100,100)[]{$(a.17)\ S=2$}

\ArrowLine(0,120)(20,120) \ArrowLine(20,120)(40,120)
\ArrowLine(40,120)(60,120) \ArrowLine(60,120)(80,120)
\PhotonArc(30,120)(10,0,180){2}{5} \Photon(60,120)(60,140){2}{3}
\ArrowArc(30,150)(10,0,180) \ArrowArc(30,150)(10,-180,0)
\Photon(20,150)(40,150){2}{3} \ArrowLine(60,140)(80,140)
\ArrowLine(80,155)(60,140) \Text(0,110)[]{$\mu$}
\Text(45,160)[]{$\mu$} \Text(80,110)[]{$\nu_\mu$}
\Text(90,160)[]{$\widetilde{\nu_e}$} \Text(90,140)[]{$e$}
\Text(40,100)[]{$(b.17)\ S=1$ }

\ArrowLine(130,120)(180,120) \ArrowLine(180,120)(195,120)
\PhotonArc(205,120)(10,0,180){2}{5} \ArrowLine(195,120)(215,120)
\ArrowLine(215,120)(230,120) \ArrowArc(155,140)(10,0,180)
\ArrowArc(155,140)(10,-180,0) \Photon(180,120)(180,140){2}{3}
\Photon(145,140)(165,140){2}{3} \ArrowLine(180,140)(230,140)
\ArrowLine(230,165)(180,140) \Text(135,110)[]{$\mu$}
\Text(165,160)[]{$\mu$} \Text(230,110)[]{$\nu_\mu$}
\Text(240,170)[]{$\widetilde{\nu_e}$} \Text(240,140)[]{$e$}
\Text(180,100)[]{$(c.17)\ S=1$ }
 \end{picture}

 \begin{picture}(400,85)(-150,250)
\ArrowLine(-150,120)(-100,120) \ArrowLine(-100,120)(-50,120)
\ArrowLine(-100,145)(-85,145) \ArrowLine(-85,145)(-65,145)
\ArrowLine(-65,145)(-50,145) \Photon(-100,120)(-100,145){2}{4}
\ArrowLine(-100,145)(-50,145) \ArrowLine(-50,195)(-100,145)
\PhotonArc(-75,170)(10,-135,45){2}{5}
\PhotonArc(-75,145)(10,-180,0){2}{5} \Text(-150,110)[]{$\mu$}
\Text(-50,110)[]{$\nu_\mu$} \Text(-50,180)[]{$\widetilde{\nu_e}$}
\Text(-40,145)[]{$e$} \Text(-100,100)[]{$(a.18)\ S=1$}

\ArrowLine(0,120)(40,120) \ArrowLine(40,120)(100,120)
\Photon(40,120)(40,130){2}{2} \Photon(40,150)(40,160){2}{2}
\ArrowArc(40,140)(10,-90,90) \ArrowArc(40,140)(10,90,-90)
\ArrowLine(100,185)(40,160) \ArrowLine(40,160)(60,160)
\ArrowLine(60,160)(80,160) \ArrowLine(80,160)(100,160)
\PhotonArc(70,160)(10,-180,0){2}{5} \Text(0,110)[]{$\mu$}
\Text(100,110)[]{$\nu_\mu$} \Text(90,190)[]{$\widetilde{\nu_e}$}
\Text(110,160)[]{$e$} \Text(50,100)[]{$(b.18)\ S=1$ }
\Text(57,140)[]{$e$}

\ArrowLine(130,120)(190,120) \ArrowLine(190,120)(240,120)
\ArrowArc(190,140)(10,-90,90) \ArrowArc(190,140)(10,90,-90)
\Photon(190,120)(190,130){2}{2} \Photon(190,150)(190,160){2}{2}
\ArrowLine(190,160)(240,160) \ArrowLine(230,200)(190,160)
\PhotonArc(210,180)(10,-135,45){2}{5} \Text(140,110)[]{$\mu$}
\Text(240,110)[]{$\nu_\mu$} \Text(240,190)[]{$\widetilde{\nu_e}$}
\Text(250,160)[]{$e$} \Text(200,100)[]{$(c.18)\ S=1$ }
\Text(210,140)[]{$\mu$}
 \end{picture}
 \ec
\end{figure}
\newpage

\begin{figure}[ht]

 \begin{picture}(400,90)(-150,100)
\ArrowLine(-150,120)(-70,120) \ArrowLine(-70,120)(-50,120)
\ArrowArc(-125,140)(10,0,180) \ArrowArc(-125,140)(10,-180,0)
\Photon(-135,140)(-115,140){2}{3} \ArrowArc(-95,140)(10,0,180)
\ArrowArc(-95,140)(10,-180,0) \Photon(-105,140)(-85,140){2}{3}
\Photon(-70,120)(-70,145){2}{4} \ArrowLine(-70,145)(-50,145)
\ArrowLine(-50,165)(-70,145) \Text(-95,160)[]{$e$}
\Text(-125,160)[]{$e$} \Text(-150,110)[]{$\mu$}
\Text(-50,110)[]{$\nu_\mu$} \Text(-50,170)[]{$\widetilde{\nu_e}$}
\Text(-40,145)[]{$e$} \Text(-100,90)[]{$(a.19)\ g_1=1, k_1=2$}
\Text(-100,80)[]{$S=2$}

\ArrowLine(0,120)(20,120) \ArrowLine(20,120)(40,120)
\ArrowLine(40,120)(60,120) \ArrowLine(60,120)(80,120)
\PhotonArc(30,120)(10,0,180){2}{5} \Photon(60,120)(60,140){2}{3}
\ArrowArc(30,150)(10,0,180) \ArrowArc(30,150)(10,-180,0)
\Photon(20,150)(40,150){2}{3} \ArrowLine(60,140)(80,140)
\ArrowLine(80,155)(60,140) \Text(0,110)[]{$\mu$}
\Text(45,160)[]{$e$} \Text(80,110)[]{$\nu_\mu$}
\Text(90,160)[]{$\widetilde{\nu_e}$} \Text(90,140)[]{$e$}
\Text(40,90)[]{$(b.19)\ S=1$ }

\ArrowLine(130,120)(180,120) \ArrowLine(180,120)(195,120)
\PhotonArc(205,120)(10,0,180){2}{5} \ArrowLine(195,120)(215,120)
\ArrowLine(215,120)(230,120) \ArrowArc(155,140)(10,0,180)
\ArrowArc(155,140)(10,-180,0) \Photon(180,120)(180,140){2}{3}
\Photon(145,140)(165,140){2}{3} \ArrowLine(180,140)(230,140)
\ArrowLine(230,165)(180,140) \Text(135,110)[]{$\mu$}
\Text(165,160)[]{$e$} \Text(230,110)[]{$\nu_\mu$}
\Text(240,170)[]{$\widetilde{\nu_e}$} \Text(240,140)[]{$e$}
\Text(180,90)[]{$(c.19)\ S=1$ }
 \end{picture}

 \begin{picture}(400,90)(-150,160)
\ArrowLine(-150,120)(-70,120) \ArrowLine(-70,120)(-50,120)
\ArrowArc(-125,140)(10,0,180) \ArrowArc(-125,140)(10,-180,0)
\Photon(-135,140)(-115,140){2}{3} \ArrowArc(-95,140)(10,0,180)
\ArrowArc(-95,140)(10,-180,0) \Photon(-105,140)(-85,140){2}{3}
\Photon(-70,120)(-70,145){2}{4} \ArrowLine(-70,145)(-50,145)
\ArrowLine(-50,165)(-70,145) \Text(-95,160)[]{$e$}
\Text(-125,160)[]{$\mu$} \Text(-150,110)[]{$\mu$}
\Text(-50,110)[]{$\nu_\mu$} \Text(-50,170)[]{$\widetilde{\nu_e}$}
\Text(-40,145)[]{$e$}\Text(-100,90)[]{$(a.20)\ g_1=g_2=1$}
\Text(-100,70)[]{$s=1$}\Text(-85,80)[]{$k_1=k_2=1$}

 \ArrowLine(-10,120)(70,120) \ArrowLine(70,120)(90,120)
\ArrowArc(15,140)(10,0,180) \ArrowArc(15,140)(10,-180,0)
\Photon(5,140)(25,140){2}{4} \ArrowArc(45,140)(10,0,180)
\ArrowArc(45,140)(10,-180,0) \Photon(35,140)(55,140){2}{4}
\ArrowArc(45,170)(10,0,180) \ArrowArc(45,170)(10,-180,0)
\Photon(55,170)(35,170){2}{4} \Photon(70,120)(70,145){2}{4}
\ArrowLine(70,145)(90,145) \ArrowLine(90,165)(70,145)
\Text(60,150)[]{$\mu$} \Text(-5,150)[]{$\mu$}
\Text(30,170)[]{$\mu$} \Text(-10,110)[]{$\mu$}
\Text(90,110)[]{$\nu_\mu$} \Text(90,170)[]{$\widetilde{\nu_e}$}
\Text(100,145)[]{$e$} \Text(45,80)[]{$S=3!=6$}
 \Text(40,90)[]{$(b.20)\ g_1=1,k_1=3$}

\ArrowLine(150,120)(210,120) \ArrowLine(210,120)(250,120)
\Photon(210,120)(210,140){2}{3} \ArrowArc(165,140)(10,0,180)
\ArrowArc(165,140)(10,-180,0) \Photon(155,140)(175,140){2}{3}
\ArrowArc(195,140)(10,0,180) \ArrowArc(195,140)(10,-180,0)
\Photon(185,140)(205,140){2}{3} \ArrowArc(195,170)(10,0,180)
\ArrowArc(195,170)(10,-180,0) \Photon(185,170)(205,170){2}{3}
\ArrowLine(210,140)(250,140) \ArrowLine(250,160)(210,140)
\Text(150,110)[]{$\mu$} \Text(160,155)[]{$e$}
\Text(205,155)[]{$e$} \Text(180,180)[]{$e$}
\Text(250,110)[]{$\nu_\mu$} \Text(260,170)[]{$\widetilde{\nu_e}$}
\Text(260,140)[]{$e$} \Text(190,80)[]{$S =6$
}\Text(190,90)[]{$(c.20)\ g_1=1,k_1=3$ }

\end{picture}

\begin{picture}(400,90)(-150,200)
\ArrowLine(-150,120)(-70,120) \ArrowLine(-70,120)(-50,120)
\ArrowArc(-125,140)(10,0,180) \ArrowArc(-125,140)(10,-180,0)
\Photon(-135,140)(-115,140){2}{3} \ArrowArc(-95,140)(10,0,180)
\ArrowArc(-95,140)(10,-180,0) \Photon(-105,140)(-85,140){2}{3}
\ArrowArc(-95,170)(10,0,180) \ArrowArc(-95,170)(10,-180,0)
\Photon(-85,170)(-105,170){2}{4} \Photon(-70,120)(-70,145){2}{4}
\ArrowLine(-70,145)(-50,145) \ArrowLine(-50,165)(-70,145)
\Text(-80,140)[]{$\mu$} \Text(-145,140)[]{$e$}
\Text(-110,170)[]{$e$} \Text(-150,110)[]{$\mu$}
\Text(-50,110)[]{$\nu_\mu$} \Text(-50,170)[]{$\widetilde{\nu_e}$}
\Text(-40,145)[]{$e$} \Text(-100,90)[]{$(a.21) \
g_1=g_2=1$}\Text(-85,80)[]{$k_1=2,k_2=1$} \Text(-100,70)[]{$S=2$}
 \ArrowLine(0,120)(20,120)
\ArrowLine(20,120)(40,120) \ArrowLine(40,120)(60,120)
\ArrowLine(60,120)(80,120) \PhotonArc(30,120)(10,0,180){2}{5}
\Photon(60,120)(60,140){2}{3} \ArrowArc(15,150)(10,0,180)
\ArrowArc(15,150)(10,-180,0) \Photon(5,150)(25,150){2}{3}
\ArrowArc(45,150)(10,0,180) \ArrowArc(45,150)(10,-180,0)
\Photon(35,150)(55,150){2}{3} \ArrowLine(60,140)(80,140)
\ArrowLine(80,155)(60,140) \Text(0,110)[]{$\mu$}
\Text(15,170)[]{$\mu$} \Text(45,170)[]{$e$}
\Text(80,110)[]{$\nu_\mu$} \Text(90,160)[]{$\widetilde{\nu_e}$}
\Text(90,140)[]{$e$} \Text(40,90)[]{$(b.21) \ S=1$ }

\ArrowLine(130,120)(180,120) \ArrowLine(180,120)(195,120)
\PhotonArc(205,120)(10,0,180){2}{5} \ArrowLine(195,120)(215,120)
\ArrowLine(215,120)(230,120) \ArrowArc(140,140)(10,0,180)
\ArrowArc(140,140)(10,-180,0) \Photon(130,140)(150,140){2}{3}
\ArrowArc(165,140)(10,0,180) \ArrowArc(165,140)(10,-180,0)
\Photon(155,140)(175,140){2}{3} \Photon(180,120)(180,140){2}{3}
\ArrowLine(180,140)(230,140) \ArrowLine(230,165)(180,140)
\Text(130,110)[]{$\mu$} \Text(165,160)[]{$\mu$}
\Text(140,160)[]{$e$} \Text(230,110)[]{$\nu_\mu$}
\Text(240,170)[]{$\widetilde{\nu_e}$} \Text(240,140)[]{$e$}
\Text(180,90)[]{$(c.21)\ S=1$ }
 \end{picture}

\begin{picture}(400,90)(-150,250)
\ArrowLine(-150,120)(-70,120) \ArrowLine(-70,120)(-50,120)
\ArrowArc(-125,140)(10,0,180) \ArrowArc(-125,140)(10,-180,0)
\Photon(-135,140)(-115,140){2}{3} \ArrowArc(-95,140)(10,0,180)
\ArrowArc(-95,140)(10,-180,0) \Photon(-105,140)(-85,140){2}{3}
\ArrowArc(-95,170)(10,0,180) \ArrowArc(-95,170)(10,-180,0)
\Photon(-85,170)(-105,170){2}{4} \Photon(-70,120)(-70,145){2}{4}
\ArrowLine(-70,145)(-50,145) \ArrowLine(-50,165)(-70,145)
\Text(-80,140)[]{$\mu$} \Text(-145,140)[]{$\mu$}
\Text(-110,170)[]{$e$} \Text(-150,110)[]{$\mu$}
\Text(-50,110)[]{$\nu_\mu$} \Text(-50,170)[]{$\widetilde{\nu_e}$}
\Text(-40,145)[]{$e$} \Text(-100,100)[]{$(a.22)\ S=2$}

\ArrowLine(0,120)(60,120) \ArrowLine(60,120)(100,120)

\Photon(60,120)(60,140){2}{3} \ArrowArc(15,140)(10,0,180)
\ArrowArc(15,140)(10,-180,0) \Photon(5,140)(25,140){2}{3}
\ArrowArc(45,140)(10,0,180) \ArrowArc(45,140)(10,-180,0)
\Photon(35,140)(55,140){2}{3} \ArrowLine(60,140)(100,140)
\ArrowLine(100,180)(60,140) \PhotonArc(80,160)(10,-135,45){2}{5}
\Text(0,110)[]{$\mu$} \Text(15,160)[]{$\mu$} \Text(45,160)[]{$e$}
\Text(100,110)[]{$\nu_\mu$} \Text(110,190)[]{$\widetilde{\nu_e}$}
\Text(110,140)[]{$e$} \Text(40,100)[]{$(b.22)\ S=1$ }

\ArrowLine(130,120)(180,120) \ArrowLine(180,120)(230,120)
\ArrowArc(140,140)(10,0,180) \ArrowArc(140,140)(10,-180,0)
\Photon(130,140)(150,140){2}{3} \ArrowArc(165,140)(10,0,180)
\ArrowArc(165,140)(10,-180,0) \Photon(155,140)(175,140){2}{3}
\Photon(180,120)(180,140){2}{3} \ArrowLine(180,140)(195,140)
\PhotonArc(205,140)(10,-180,0){2}{5} \ArrowLine(195,140)(215,140)
\ArrowLine(215,140)(230,140) \ArrowLine(230,165)(180,140)
\Text(130,110)[]{$\mu$} \Text(165,160)[]{$\mu$}
\Text(140,160)[]{$e$} \Text(230,110)[]{$\nu_\mu$}
\Text(240,170)[]{$\widetilde{\nu_e}$} \Text(240,140)[]{$e$}
\Text(180,100)[]{$(c.22)\ S=1$ }
 \end{picture}
 \begin{picture}(400,90)(-150,280)
\ArrowLine(-150,120)(-70,120) \ArrowLine(-70,120)(-50,120)
\ArrowArc(-125,140)(10,0,180) \ArrowArc(-125,140)(10,-180,0)
\Photon(-135,140)(-115,140){2}{3} \ArrowArc(-125,170)(10,0,180)
\ArrowArc(-125,170)(10,-180,0) \Photon(-135,170)(-115,170){2}{3}
\ArrowArc(-95,140)(10,0,180) \ArrowArc(-95,140)(10,-180,0)
\Photon(-105,140)(-85,140){2}{3} \ArrowArc(-95,170)(10,0,180)
\ArrowArc(-95,170)(10,-180,0) \Photon(-85,170)(-105,170){2}{4}
\Photon(-70,120)(-70,145){2}{4} \ArrowLine(-70,145)(-50,145)
\ArrowLine(-50,165)(-70,145) \Text(-80,140)[]{$\mu$}
\Text(-80,170)[]{$e$} \Text(-145,140)[]{$\mu$}
\Text(-145,170)[]{$e$} \Text(-150,110)[]{$\mu$}
\Text(-50,110)[]{$\nu_\mu$} \Text(-50,170)[]{$\widetilde{\nu_e}$}
\Text(-40,145)[]{$e$} \Text(-100,90)[]{$(a.23)\ g_1=g_2=1$}
\Text(-100,80)[]{$k_1=k_2=2$}\Text(-100,70)[]{$S=4$}
\ArrowLine(0,120)(60,120) \ArrowLine(60,120)(100,120)
\Photon(60,120)(60,140){2}{3} \ArrowArc(15,140)(10,0,180)
\ArrowArc(15,140)(10,-180,0) \Photon(5,140)(25,140){2}{3}
\ArrowArc(15,170)(10,0,180) \ArrowArc(15,170)(10,-180,0)
\Photon(5,170)(25,170){2}{3} \ArrowArc(45,140)(10,0,180)
\ArrowArc(45,140)(10,-180,0) \Photon(35,140)(55,140){2}{3}
\ArrowArc(45,170)(10,0,180) \ArrowArc(45,170)(10,-180,0)
\Photon(35,170)(55,170){2}{3} \ArrowLine(60,140)(100,140)
\ArrowLine(100,160)(60,140)
\Text(0,110)[]{$\mu$} \Text(0,145)[]{$e$} \Text(55,150)[]{$e$}
\Text(55,180)[]{$e$} \Text(0,180)[]{$\mu$}
\Text(100,110)[]{$\nu_\mu$} \Text(110,170)[]{$\widetilde{\nu_e}$}
\Text(110,140)[]{$e$} \Text(40,90)[]{$(b.23)\ g_1=g_2=1$ }
\Text(40,80)[]{$k_1=3, k_2=1$ }\Text(40,70)[]{$S =6$ }

\ArrowLine(130,120)(200,120) \ArrowLine(200,120)(230,120)
\ArrowArc(150,140)(10,0,180) \ArrowArc(150,140)(10,-180,0)
\Photon(140,140)(160,140){2}{3} \ArrowArc(150,170)(10,0,180)
\ArrowArc(150,170)(10,-180,0) \Photon(140,170)(160,170){2}{3}
\ArrowArc(180,140)(10,0,180) \ArrowArc(180,140)(10,-180,0)
\Photon(170,140)(190,140){2}{3} \ArrowArc(180,170)(10,0,180)
\ArrowArc(180,170)(10,-180,0) \Photon(170,170)(190,170){2}{3}
\Photon(200,120)(200,140){2}{3} \ArrowLine(200,140)(230,140)
\ArrowLine(230,165)(200,140) \Text(130,110)[]{$\mu$}
\Text(195,150)[]{$e$} \Text(140,150)[]{$e$} \Text(195,180)[]{$e$}
\Text(140,180)[]{$e$} \Text(230,110)[]{$\nu_\mu$}
\Text(250,170)[]{$\widetilde{\nu_e}$} \Text(240,140)[]{$e$}
\Text(180,90)[]{$(c.23) \ g_1=1,k_1=4$ }\Text(180,80)[]{$S=24$ }
 \end{picture}

\end{figure}

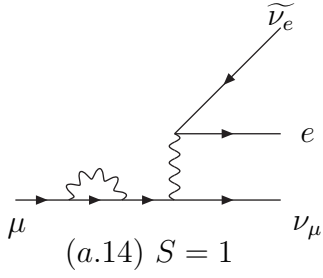
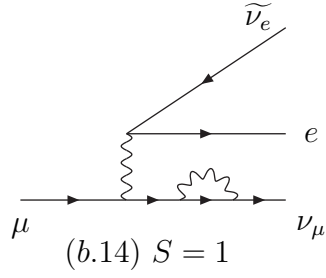
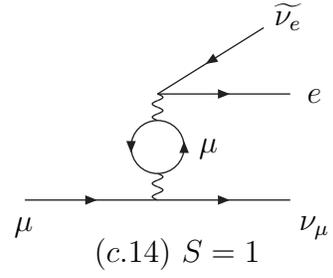
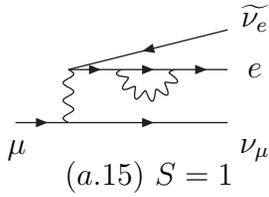
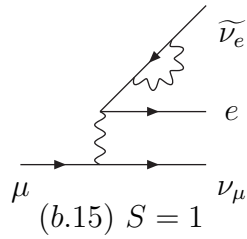
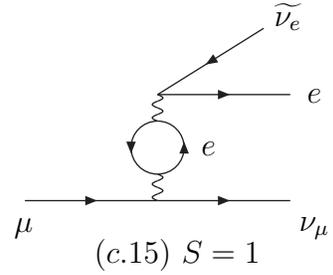
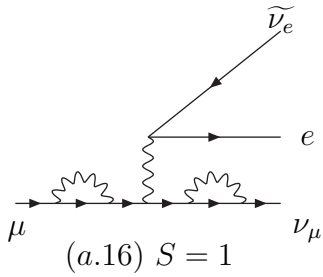
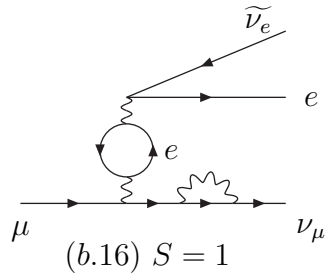
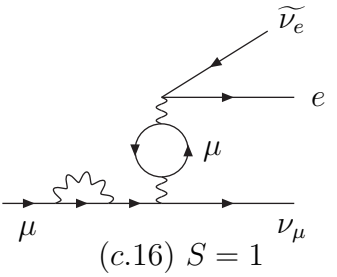
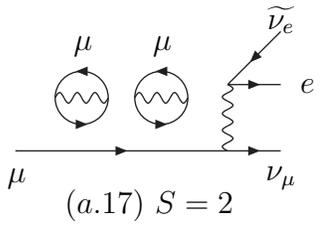
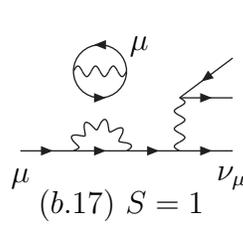
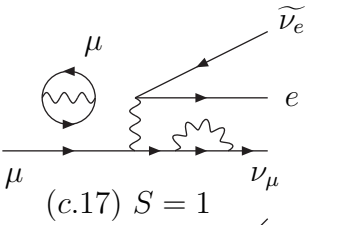
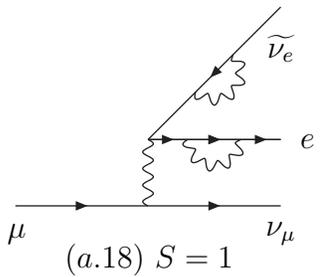
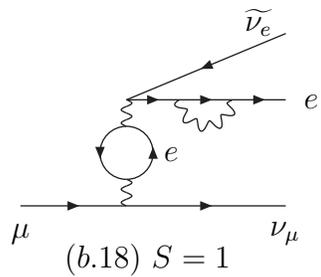
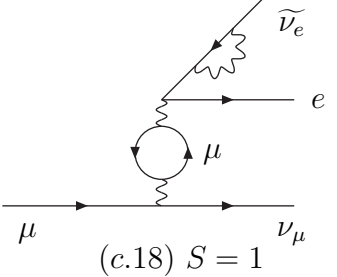
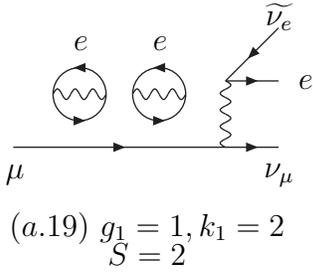
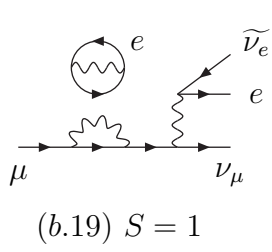
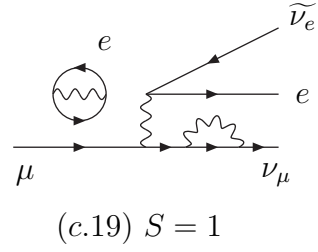
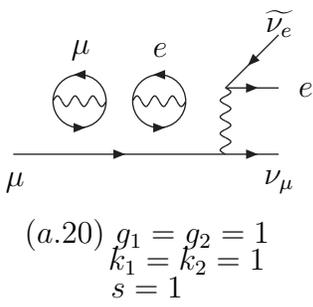
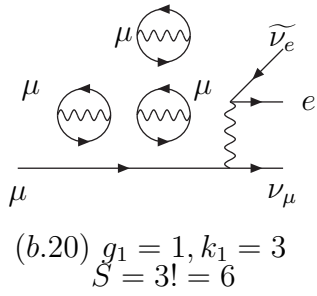
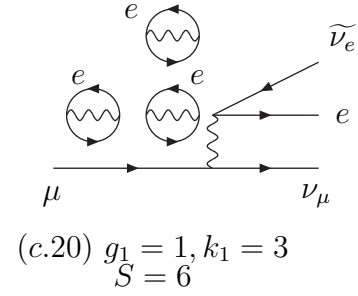
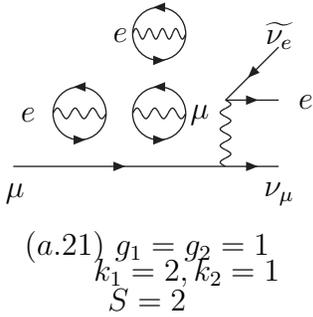
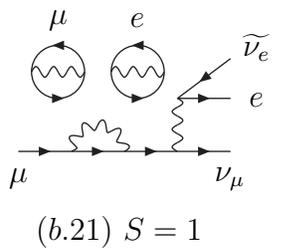
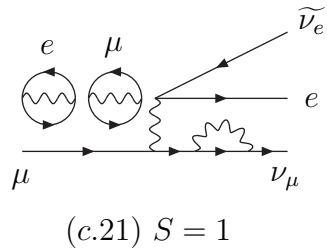
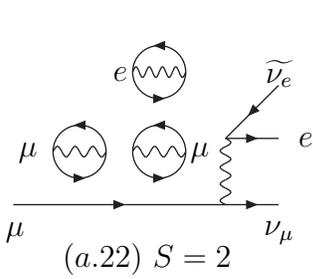
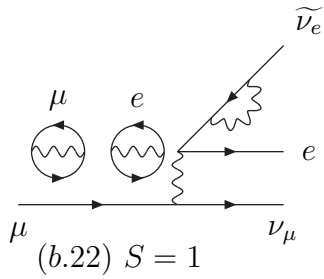
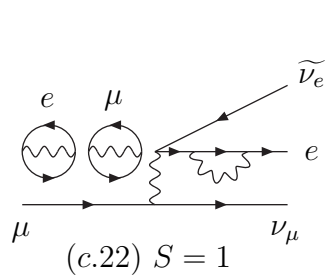
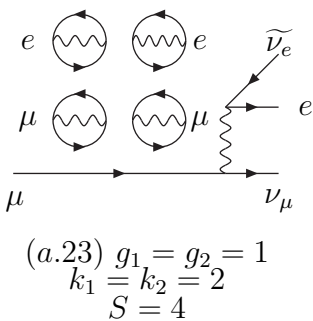
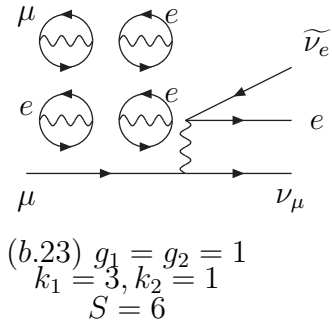
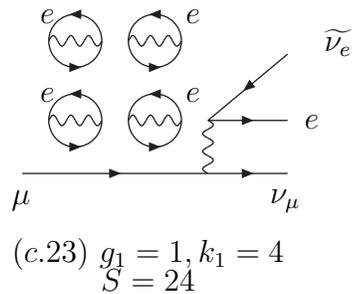
\end{document}